\documentclass[sigconf]{acmart}
\usepackage{etex}
\usepackage{subcaption} %% For complex figures with subfigures/subcaptions
\let\terms\undefined

\usepackage{booktabs}   %% For formal tables:
\usepackage{subcaption} %% For complex figures with subfigures/subcaptions
\usepackage[noend]{algpseudocode}

\usepackage{graphicx}
\usepackage{amsmath}
\usepackage{xspace}
\usepackage{footnote}
\usepackage{amsfonts}
\usepackage{hhline}
\usepackage{multirow}
\usepackage{setspace} 
\usepackage{epsfig}
\usepackage{url}
\usepackage{booktabs}

\usepackage{xcolor}
\usepackage{lipsum}
\usepackage{courier}
\usepackage{listings}
\usepackage{caption}
\usepackage{colortbl}
\usepackage{arydshln} %dashed lines
\usepackage{makecell} %format table cells
\usepackage{tikz}
\usepackage{xcolor}
\usetikzlibrary{calc}
\usepackage{varwidth}
\usepackage{natbib}
\usepackage{algorithm}
\usepackage[noend]{algpseudocode}
\usepackage{subcaption}

\usetikzlibrary{fit,arrows,calc,positioning,quotes}
\usetikzlibrary{decorations.pathreplacing}

\usepackage{mathtools}
\usepackage{fancyvrb}
\usepackage{textcomp}
\usepackage{stmaryrd}

\tikzset{
  -|-/.style={
    to path={
      (\tikztostart) -| ($(\tikztostart)!#1!(\tikztotarget)$) |- (\tikztotarget)
      \tikztonodes
    }
  },
  -|-/.default=0.5,
  |-|/.style={
    to path={
      (\tikztostart) |- ($(\tikztostart)!#1!(\tikztotarget)$) -| (\tikztotarget)
      \tikztonodes
    }
  },
  |-|/.default=0.5
}

\usetikzlibrary{shapes.geometric}

\long\def\com#1{}
\newcommand{\app}{Clef\xspace}

\newcommand{\betterthangroundtruth}{\ensuremath{40.6\%}\xspace}
\newcommand{\overallaccuracy}{\ensuremath{42.1\%}\xspace}
\newcommand{\hasrepairaccuracy}{\ensuremath{50.0\%}\xspace}
\newcommand{\pureincorrectaccuracy}{\ensuremath{34.1\%}\xspace}

\newcommand{\abandonrate}{\ensuremath{52.2\%}\xspace}

\newcommand{\para}[1]{\smallskip\noindent {\bf #1}}

\newcommand{\squishlist}{
   \begin{list}{$\bullet$}
    { \setlength{\itemsep}{0pt}      \setlength{\parsep}{3pt}
      \setlength{\topsep}{3pt}       \setlength{\partopsep}{0pt}
      \setlength{\leftmargin}{3.5mm} \setlength{\labelwidth}{1em}
      \setlength{\labelsep}{0.5em} } 
}

\newcommand{\squishend}{
    \end{list}  }

\def\BibTeX{{\rm B\kern-.05em{\sc i\kern-.025em b}\kern-.08em
    T\kern-.1667em\lower.7ex\hbox{E}\kern-.125emX}}

\begin{document}

\title{Automated Feedback Generation for Competition-Level Code}

\author{Jialu Zhang}
\affiliation{%
  \institution{Yale University}
  \city{New Haven}
  \state{Connecticut}
  \postcode{06511}
  \country{USA}
  }
  
\author{De Li}
\affiliation{%
  \institution{The MathWorks, Inc.}
  \city{Natick}
  \state{Massachusetts}
  \postcode{01760}
  \country{USA}
  }

\author{John C. Kolesar}
\affiliation{%
  \institution{Yale University}
  \city{New Haven}
  \state{Connecticut}
  \postcode{06511}
  \country{USA}
  }
  
\author{Hanyuan Shi}
\affiliation{%
  \institution{Independent Researcher}
  \city{Hangzhou}
  \state{Zhejiang}
  \postcode{ }
  \country{China}
  }

\author{Ruzica Piskac}
\affiliation{%
  \institution{Yale University}
  \city{New Haven}
  \state{Connecticut}
  \postcode{06511}
  \country{USA}
  }

\begin{abstract}
	Competitive programming has become a popular way for programmers to test their skills.
Large-scale online programming contests attract millions of experienced programmers to compete against each other.
Competition-level programming problems are challenging in nature, and participants often fail to solve the problem on their first attempt.
%Generally, most participants in a programming contest fail to produce a correct submission because of the complexity of the problems and the rigorous evaluation process.
Some online platforms for competitive programming allow programmers to practice on competition-level problems as well, and the standard feedback for an incorrect practice submission is the first test case that the submission fails.
%Generally, programming competition platforms provide feedback of the same form for programmers practicing on competition-level problems.
%The current standard feedback for programmers who submit an incorrect program while practicing for a competition is the first test case that the submission fails.
Often, the failed test case does not provide programmers with enough information to resolve the errors in their code, and they abandon the problem after several more unsuccessful attempts.

We present \app, the first data-driven tool that can generate feedback on competition-level code automatically by repairing programmers' incorrect submissions.
The key development is that \app can learn how to generate repairs for incorrect submissions by examining the repairs that other programmers made to their own submissions over time.
Since the differences between an incorrect program and a correct program for the same task may be significant, we introduce a new data structure, \textit{merge trees}, to capture the changes between submissions.
Merge trees are versatile: they can encode both large algorithm-level redesigns and small statement-level alterations.
\app applies the patterns it learns from a database of submissions to generate repairs for new submissions outside the database.
We evaluated \app on six real-world problems from Codeforces, the world's largest platform for competitive programming.  \app achieves \overallaccuracy accuracy in repairing programmers' incorrect submissions.
Even when given incorrect submissions from programmers who never found the solution to a problem on their own, \app repairs the users' programs \pureincorrectaccuracy of the time.

\if
The educational value of programming assignments depends critically on instructors' ability to deliver timely feedback for students' submissions, but repairing buggy submissions manually is an expensive and onerous job.  For large-scale online programming contests, where submissions can number in the thousands, providing feedback manually is a completely infeasible task.

Existing techniques for automated program repair suffer from a range of problems. Some needs users to provide an inputs as the guidance to the tool, thus have low automation level. The other support some specific error types or focusing on repairing introductory programming assignments, thus fail to generate complex repairs.  We propose a new automated program repair technique for programming contest assignments.  Our key idea is to learn how existing users repair their own submissions and learn program transformation patterns to repair target student submissions.

We evaluated our tool, \app, on our training corpus, which contains more than a hundred thousand real-world submissions from students. 
\fi

\end{abstract}

\maketitle

\section{Introduction}
\label{sec:intro}
Competitive programming enjoys widespread popularity.
The International Collegiate Programming Contest (ICPC), one of the most prestigious programming contests for college students, has been held annually for more than 50 years.
Each year, more than 50,000 students from over 3,000 universities in over 100 countries compete for medals in the contest~\cite{ICPC}.
Moreover, competitive programming has had a significant impact in industry as well.
Platforms such as Codeforces\footnote{https://codeforces.com} and Topcoder\footnote{https://www.topcoder.com} host large-scale online programming contests that attract millions of experienced programmers.
Software companies view the finalists in the competitions as strong candidates for hiring since the finalists demonstrate solid algorithmic problem-solving skills and an outstanding ability to handle on-the-spot stress.
Some software companies, such as Google~\cite{companyContest}, Meta~\cite{FacebookHacker}, Microsoft~\cite{microsoftImagineCup}, Yandex~\cite{Yandexalgo}, and HP~\cite{hpecodewars}, even hold their own online programming contests for recruiting purposes~\cite{companyContest}.

%However, competitive programming is also very challenging.
In a programming competition, participants receive a description of a problem and a short list of sample tests illustrating
how the program should execute on some example inputs.
The participants develop solutions for the problem 
and submit them. The solutions are evaluated automatically on a number of different tests that are hidden from the participants.
If a solution passes every test in the suite, it is accepted as correct.
Competition-level problems are non-trivial:  correct implementations sometimes require hundreds of lines of code, and the entire program needs to be efficient and bug-free.
Additionally, the platform for the competition evaluates programmers' submissions rigorously on multiple criteria.
The automatic tests can involve carefully-designed corner cases, along with hard limits on execution time and memory usage.

\para{State-of-the-art feedback generation for programmers' incorrect submissions.}
Some competitive programming platforms allow programmers to practice on problems from past competitions.
Currently, the standard response to a programmer's incorrect practice submission is simply to expose the first test case that the submission fails.
Even with test case exposure for failures, many programmers still fail to solve the problem in the end, and they abandon the problem after several unsuccessful submissions.
For the problems that we surveyed, \abandonrate of the incorrect submissions were never corrected fully by their authors.
Forums can serve as a helpful source of feedback for programmers while they are practicing, but users who post questions on a forum have no guarantee of a timely response.
%but providing feedback manually is unfeasible for large programming contests since the participants can number in the tens of thousands.
%the core of the algorithm for solving a competitive programming problem is often about inventing an efficient algorithm, the availability of this type of manual feedback becomes an issue. 
%The organizers of a large programming contest cannot hope to provide manual feedback for every incorrect submission in a timely fashion, and whatever manual feedback they do provide may be incomplete or even incorrect.
Furthermore, the feedback from other users on the forum can be incomplete or incorrect, and the users who post the questions might not ask for the right information in the first place because they misunderstand their own problems.
A tool that repairs programs or provides useful feedback automatically would be a helpful alternative.
%Having a tool that could automatically repair an incorrect program or provide useful feedback would be very helpful 

In recent years, researchers in the automated program repair community have worked on generating feedback automatically for intro-level programming assignments ~\cite{Autograder,Refazer,clara,SAR,Refactory,Cafe}. %but their focus are intro-level problems
State-of-the-art feedback generators use data-driven approaches.
They all take advantage of databases of previously-written correct submissions to learn how to repair new incorrect submissions.
Unfortunately, these tools target only problems from intro-level programming courses, and their 
feedback generation techniques do not suffice for competition-level problems.
Two major differences exist between intro-level and competition-level programming:
\begin{itemize}
    \item \textbf{The difficulty level of the problems.} Intro-level programming problems focus primarily on training programmers to use the features of a language correctly~\cite{APPS}.
    Consequently, they usually do not require programmers to use a specific algorithm or data structure.
    Typical intro-level problems include printing a chessboard~\cite{SAR} and computing the derivative of a function~\cite{clara}.
    On the other hand, competition-level programming problems require programmers to understand complex natural-language descriptions, to master a wide range of algorithms and data structures, and to implement solutions that may involve hundreds of lines of code.
    
    \item \textbf{The evaluation metrics.} Intro-level programming problems usually do not have rigorous evaluation metrics.
    Restrictions on execution time and memory consumption are rare:  generally, the test suites for intro-level problems only cover functional correctness.
    On the other hand, evaluation suites for competition-level problems perform rigorous checks on execution time and memory consumption.
    Any timeout or excessive memory usage causes the suite to mark a submission as incorrect even if the program behaves perfectly in terms of functional correctness.
    
    %\item \textbf{The different usage of test cases.} Compared to intro-level programming problems, there are fewer sample test cases given and more unseen, corner cases in testing phase in the competitive programming. 
    
\end{itemize}

Furthermore, state-of-the-art intro-level feedback generators suffer from a variety of weaknesses, including the inability to generate complex repairs~\cite{Refazer,SAR,Refactory}, the tendency to enlarge programs excessively with repairs~\cite{clara}, and dependence on manual guidance from users~\cite{Autograder}.
Finding an effective automatic feedback generation method for competition-level code remains an open problem.

Problem 1475A from Codeforces illustrates some of the major differences between intro-level and competition-level problems.
%The following example illustrates some of the major differences. It is problem 1475A on Codeforces.
The input is an integer $n$ $(2 < n < 10^{14})$, and the goal is to determine whether $n$ has any odd divisor greater than one.\footnote{https://codeforces.com/contest/1475/problem/A}
The execution time limit for Problem 1475A is two seconds per test, and the memory limit is 256 MB per test.
%A user also ask for help in solving this problem on Stack Overflow~\cite{oddDivisorStackoverflow}.
One solution for the problem posted on Stack Overflow~\cite{oddDivisorStackoverflow} performs an exhaustive search by iterating over every odd number in $(1, n)$ and checking whether $n$ is divisible by it:

\begin{lstlisting}
for (unsigned long long i=3; i<n; i+=2){
   if (n%i == 0) return true; //It has an odd divisor
}
return false; // n%i == 0 was never true so it doesn't have an odd divisor
\end{lstlisting}

This submission is syntactically and semantically correct.
However, it fails to pass the evaluation suite:  the test suite marks it as incorrect with ``Time limit exceeded on test 2'' as the feedback.
Solving the problem correctly within the time limit requires a more efficient algorithm, and finding that efficient algorithm requires an important insight:  the odd divisor problem reduces to checking whether $n$ is a power of two.
An efficient program for Problem 1475A right-shifts $n$ repeatedly to remove trailing zeroes and then checks at the end that the remaining one is the only one in $n$:

\begin{lstlisting}
while (!(n&1)) n >>= 1;
if (n==1) return false; else return true;
\end{lstlisting}

Problem 1475A presents a challenge for automated feedback generation tools.  A submission that approaches the problem incorrectly may require a complete algorithm-level redesign, not just a small local repair.
The automatic feedback provided by Codeforces did not help the programmer who wrote the exhaustive-search implementation to see that a completely different approach was necessary.
Additionally, state-of-the-art tools~\cite{Autograder,Refazer,clara,SAR,Refactory} cannot make the repairs that the program needs because they view correctness only in terms of input-output correspondence, not efficiency.

\para{Our approach: \app.}
%Our paper answers a key research question: \textbf{how to help programmers when their programs are incorrect?}
We introduce CLEF (Competition-Level Effective Feedback), a new tool that generates feedback automatically for competition-level problems by repairing errors in programmers' incorrect submissions.
\app learns the repair patterns that it uses by analyzing the repairs that other programmers made for the same problem across their submission histories.
%These patterns include the algorithm-level code redesign to statement-level running repairs.
\app applies the patterns that it learns to target programs outside the database to generate candidate repaired programs.
%Additionally, \app performs variable usage analysis to eliminate spurious candidate programs.

\para{Main technical challenges in designing \app.}
The main technical challenge for \app is having an effective method for learning how programmers repair errors in their own submissions.
The repair patterns that \app needs to learn range from small statement-level fixes to algorithm-level control flow redesigns.
%extending this way significantly the numbers of submissions that can be automatically repaired.
Other data-driven feedback generation tools cannot alter the control flow of an incorrect program~\cite{SAR,clara}, so large-scale algorithm-level changes, precisely the kind of changes that incorrect submissions for competition-level problems often require, are impossible for them to make.

\app employs a number of techniques that no other feedback generation tool has used previously:
\begin{itemize}
\item We introduce \textit{merge trees}, a new data structure for encoding both statement-level and algorithm-level changes between incorrect programs and corrected versions of the same programs.
%\item We proposed a customized matching and repairing algorithm to identify the similar pattern in the unseen incorrect submission. To this way, \app was able to repair incorrect submissions even though the errors are not exactly the same as the knowledge base.
\item We propose a new matching and repairing algorithm that takes advantage of similarities between the target program and programs in the database.  With the new algorithm, \app can repair incorrect submissions even if the errors in the submission have no exact matches in the database.
%\item \app is a self-learning system. We augmented the buggy program and the repaired program as a pair in the training set, ---this way we automatically taught \app by introducing more incorrect-remedy code patterns in the enlarged training set to improve the fixrate and repair quality for future incorrect submission.
\iffalse
\item \app learns how to improve itself.  Whenever we repair an incorrect input program, we add the original incorrect program and the repaired version of the program to the training set.  By expanding its own training set, \app improves its coverage of different coding errors, thereby improving its success rate and repair quality for future submissions for the same problem.
\fi
%\item \app is a self-learning system. For the submission that has the error that have not appeared in any previous submissions, \app use a probabilistic method, replacing the most suspicious AST with the AST selected from the correct submission. If the new candidate program is correct, we augmented the buggy program and the repaired program as a pair in the training set, ---this way we automatically taught \app by introducing more incorrect-remedy code patterns in the enlarged training set.

\end{itemize}

\para{Evaluation.} To evaluate our tool, we run \app on thousands of submissions for six real-world competitive programming problems obtained from Codeforces.
%Codeforces categorizes the problems into a number of difficulty levels.
\app provides feedback successfully for \overallaccuracy of the incorrect solutions overall.
Whenever the database contains both incorrect submissions and a correct submission for an individual user, we have \app attempt to fix the incorrect submissions without seeing the correct version, and then we compare \app's repaired version to the real user's correct version in terms of program editing distance.
In \betterthangroundtruth of the cases, \app generates a repaired program that is syntactically closer to the original incorrect submission than the user's own corrected version is.
For the cases where a user made incorrect submissions but never made a correct submission, \app repairs the user's incorrect submissions successfully \pureincorrectaccuracy of the time.

In summary, we make the following contributions:

\begin{itemize}
\item We conduct a survey to assess the characteristics and challenges of competitive programming. 
\item We present a data-driven tool, \app, that generates feedback for users' incorrect submissions automatically using its knowledge of how other users repair their own programs.
\item We propose a new data structure for capturing the algorithm-level design changes in repaired submissions.
\item We evaluate \app on real-world competitive programming problems. 
For the incorrect submissions that were later repaired by the same user, \app provides correct repairs \hasrepairaccuracy of the time. In \betterthangroundtruth of the cases, \app generates repaired programs that are closer to the original incorrect submission than the user's own correct submission.
For the incorrect submissions that were never repaired by their users, \app provides correct repairs \pureincorrectaccuracy of the time.
\end{itemize}

\section{Understanding Competitive Programming}
\label{sec:survey}

\begin{figure*}[!t]
\centering
\begin{subfigure}{.5\textwidth}
  \centering
  \includegraphics[width=.9\linewidth]{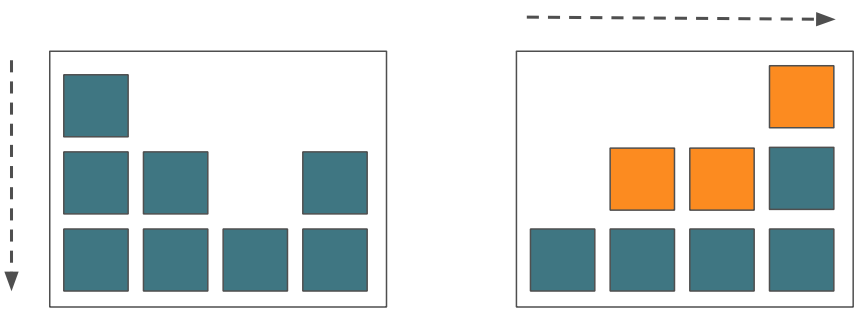}
  \caption{Original Sample Illustration}
  \label{subfig:gravity}
\end{subfigure}%
\begin{subfigure}{.5\textwidth}
  \centering
  \includegraphics[width=.9\linewidth]{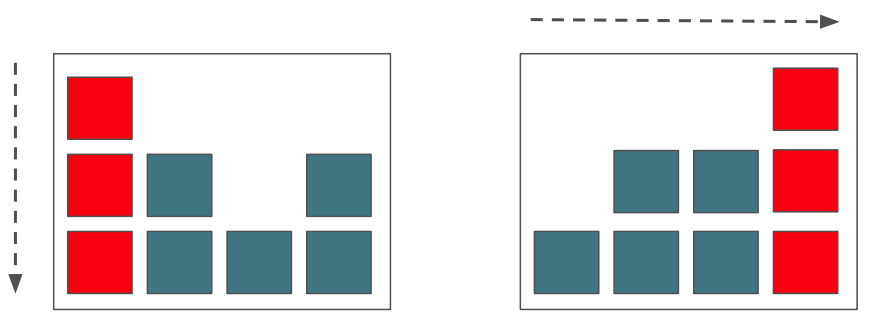}
  \caption{Required New Understanding}
  \label{subfig:gravity2}
\end{subfigure}
\caption{Subfigure~\ref{subfig:gravity} is the original sample illustration from Codeforces.
The initial configuration of the cubes in the box appears on the left, and the final configuration appears on the right.
The cubes whose positions change are highlighted in orange.
The top cube of the first column falls to the top of the last column, the top cube of the second column falls to the top of the third column, and the middle cube of the first column falls to the top of the second column.
Subfigure~\ref{subfig:gravity2} shows the same example input but highlights a different detail.
The tallest column at the end is of the same height as the tallest column at the start, but it appears at the right end of the box.
The number of columns of a given height is preserved, so the two-dimensional gravity flip problem reduces to one-dimensional array sorting.}
\label{fig:understandgravity}
\end{figure*}

%some points

% How does competitive programming look like?
% What is the most challenging part of it?
% What is the most valuable things for people to care?
% Why it is a challenges
% What is the current feedback to users? Why they are not helpful? 

In this section, we present our empirical study of real-world competitive programming.
%We describe the statement and statistics in competitive programming problems. 
We illustrate the challenges involved with solving competitive programming problems through some concrete examples. 
We also discuss the implications that drive the design of \app.

In a programming competition, a contestant writes a program to perform a specific task and submits the code to an online evaluation platform.
The platform compiles the program and runs it on a suite of pre-designed test cases.
There are seven possible outcomes for a submission:
\begin{itemize}
    \item \textbf{Accepted.} The submission produces the correct output for every test and never violates the time and memory limits.
    \item \textbf{Compile-Time Error.} The submission has at least one compilation error.  Most programs with syntax errors fall into this category.
    \item \textbf{Runtime Error.} The submission encounters an error at runtime for a test.  Common errors include buffer overflow and invalid array indices.
    \item \textbf{Time Limit Exceeded.} The program surpasses the execution time limit on a test.
    \item \textbf{Memory Limit Exceeded.} The program surpasses the memory usage limit on a test.
    \item \textbf{Wrong Answer.} The program returns an incorrect output for a test.
    \item \textbf{Other.} A non-deterministic error, such as a network outage, occurs.
\end{itemize}

%For certain programming problems, one of the three components may be especially difficult.
%We will illustrate the point with a few real-world examples.

The major sources of difficulty for competition-level problems are categorically different from the sources of difficulty for intro-level problems.
The aim of a competition-level problem's design is not to teach contestants how to write programs but to push contestants to the limits of their knowledge.
We will highlight some of the patterns in competition-level problems' designs with a few real-world examples.

\para{Pattern~1: Challenging Problem Descriptions.}
% Is there any way that we could show that the reason why users fail in this problem is because of wrong understanding of the problem.
The first step in solving a competition-level problem is converting the natural-language problem description into an idea for an algorithm.
Intro-level programming problems generally have short, straightforward descriptions, but competition-level problems can have lengthy descriptions designed to mislead contestants.
The length of a challenging problem description comes not from insignificant clutter but from complicated explanations of problem details meant to test how well programmers can bridge the gap between an end goal and an algorithm to accomplish it. For instance, consider Problem 405A from Codeforces:\footnote{https://codeforces.com/contest/405/problem/A}

\definecolor{Beau blue}{HTML}{BCD4E6}
\begin{tikzpicture}[line width=.7pt, inner sep=7pt]
  \node [draw,rounded corners,dashed,fill=Beau blue,text width=.42\textwidth] (B){%
  \bfseries 
A box contains $n$ columns of toy cubes arranged in a line.  Initially, gravity pulls all of the cubes downward, but, after the cubes are settled in place, gravity switches to pulling them to the right side of the box instead.
The input is the initial configuration of the cubes in the box, and the goal is to print the configuration of the box after gravity changes. The sample case example provided by Codeforces is shown in the left subfigure in Figure~\ref{fig:understandgravity}.};
  \node [xshift=0.1cm, yshift=0.2cm] [rotate=-30]at($(B.east)!.7!(B.north east)$) {\includegraphics[scale=.07]{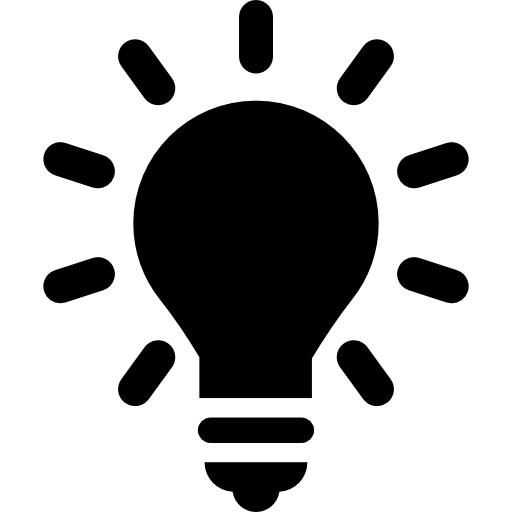}};
\end{tikzpicture}
%Many problems have complex problem description, but there are some problems that are designed to test if the programmers have the ability to reduce it to a classic algorithm.

\iffalse  
\begin{figure*}[h]
\centering\includegraphics[width=.65\linewidth]{figs/gravity_flip.png}
\caption{The initial and final configurations of the cubes in the box.  The cubes whose positions change are highlighted in orange. The top cube of the first column falls to the top of the last column, the top cube of the second column falls to the top of the third column, and the middle cube of the first column falls to the top of the second column.} 
\label{fig:gravity}
\end{figure*}
\fi

%The figure shows the initial and final configurations of the cubes in the box: the cubes that have changed their position are highlighted with orange.

The prompt of Problem 405A is designed to test programmers' ability to reduce a complex problem to a well-known simple algorithm, namely sorting.
Attempting to write a brute-force implementation that treats the cubes as distinct entities is a tedious and error-prone process.
A key insight for solving the problem is the fact that, when gravity changes, the highest columns always appear at the right end of the box and are of the same height as the highest columns at the start.
The right subfigure in Figure~\ref{fig:understandgravity} illustrates this.
The possibility of reducing the problem to sorting a one-dimensional array becomes clear after a programmer notices how the columns behave.

\iffalse
\begin{figure*}[h]
\centering\includegraphics[width=.65\linewidth]{figs/gravity_flip2.png}
\caption{The insight to reduce the two-dimensional gravity flip problem to a one-dimensional array sorting problem.} 
\label{fig:gravity2}
\end{figure*}
\fi

\para{Pattern~2: Challenging Implementation Details.}
\label{para:bacteria}
Not every single competition-level programming problem is a simple task hidden behind a complex description.
Often, implementing an effective algorithm for the problem is a genuinely difficult task involving minor details that are easy to mishandle.
Problem 579A from Codeforces\footnote{https://codeforces.com/problemset/problem/579/A} is one such problem:
%\jialu{Do you like the way to show this problem description? Like a light bulb?}

\begin{tikzpicture}[line width=.7pt, inner sep=7pt]
  \node [draw,rounded corners,dashed,fill=Beau blue,text width=.42\textwidth] (B){%
  \bfseries Start with an empty box. Each morning, you can put any number of bacteria into the box. Each night, every bacterium in the box will split into two bacteria. To get exactly $x$ ($1 \leq x \leq 10^{9}$) bacteria in the box at some moment, what is the minimum number of bacteria you need to put into the box across some number of days?};
  \node [xshift=0.1cm, yshift=0.1cm] [rotate=-30]at($(B.east)!.7!(B.north east)$) {\includegraphics[scale=.07]{figs/bulb.png}};
\end{tikzpicture}

A key insight for solving the problem is the fact that every bacterium placed in the box will become $2^n$ bacteria after $n$ days.
What the problem really requires is an algorithm that can divide one integer into a sum of powers of two.
A natural implementation for this algorithm is
%to convert $x$ from decimal to binary and then
to count the number of ones that appear in the binary representation of $x$.
Using the provided integer representation of $x$ makes this easy, but a program that converts $x$ into a binary string instead to count the number of ones can fall victim to certain errors if implemented carelessly.
%However, there is one problem that prevents programmers from using such a simple solution.
%$x$ may be a very large number, so, if the program uses a 64-bit int to store the binary representation, it will encounter an overflow error
String operations may misinterpret the base-2 string as a base-10 string, and this can lead to incorrect answers or even overflow errors.\footnote{https://stackoverflow.com/questions/52548304/converting-decimal-to-binary-with-long-integer-values-c}
To avoid overflow errors, a better program for Problem 579A never converts $x$ into an alternative format.
Instead, it operates directly on the binary representation of the integer.
It right-shifts $x$ repeatedly one bit at a time and counts the number of iterations where the right-shifted version of $x$ is odd:
\begin{lstlisting}
while (x > 0){
    if (x & 1)	r += 1; 
    x >>= 1;
}
\end{lstlisting}

\para{Pattern~3: Challenging Efficiency Requirements.}
%corner case: double? running cases?
%raising bacteria
For other problems, meeting the evaluation suite's efficiency requirements is the main source of difficulty.
Consider the following problem from GeeksforGeeks:\footnote{https://practice.geeksforgeeks.org/problems/majority-element-1587115620/1}

\begin{tikzpicture}[line width=.7pt, inner sep=7pt]
  \node [draw,rounded corners,dashed,fill=Beau blue,text width=.42\textwidth] (B){%
  \bfseries Given an array of $n$ ($1 \leq n \leq 10^{7}$) elements, find the majority element in the array, if it exists. A majority element is an element that appears more than $n/2$ times in the array.};
  \node [xshift=0.1cm, yshift=-0.05cm] [rotate=-30]at($(B.east)!.7!(B.north east)$) {\includegraphics[scale=.07]{figs/bulb.png}};
\end{tikzpicture}

The time complexity limit for the problem is $\mathcal{O}(n)$, and the space limit is $\mathcal{O}(1)$.\footnote{The stated time complexity requirement is only an indirect indicator of the real efficiency requirement used by the evaluation since the constant factors are hidden. In practice, the actual running time of the algorithm is what matters, but the time complexity is the best measurement that we can use here. The same applies for space complexity.}
The time and space requirements rule out a naive brute-force solution because a brute-force search requires two loops to keep track of the number of occurrences of every element.
Binary search and a hash map-based solution are more efficient alternatives but are still not efficient enough.
The time complexity of a binary search solution is $\mathcal{O}(n\log(n))$, and a hash map solution has a space complexity of $\mathcal{O}(n)$.

One algorithm that can solve the problem within the time and space constraints is the Boyer-Moore majority voting algorithm~\cite{Moore91}.
The algorithm starts by treating the first element of the array as the presumed majority element.  It iterates through the array, maintaining a single integer counter that starts at one.  If an element in the array is equal to the presumed majority element, increase the counter by one, and decrease the counter by one otherwise.  If the counter ever reaches zero, then reset it to one and make the element at the current index the new presumed majority element.
When the first traversal finishes, fix the presumed majority element and compare every element in the array to it.
If more than $n/2$ elements in the array are equal to the presumed majority element, the presumed majority element is the real majority element.
Otherwise, there is no majority element.
%If the counter is greater than $n/2$ when the array traversal finishes, the current presumed majority element is the real majority element.  Otherwise, there is no majority element.
%\subsection{Methodology}

%\subsection{General Findings}

%\subsection{Patterns and Examples}

% when understanding of the question is the bottleneck (bubble sort
% when precluding of unseen test cases is the bottleneck (program pass the sample test, but could not passed the 
% when the understanding to algorithm design itself is the bottle neck (odd advisor)
% when algorithm to neat implementation is the bottleneck (complexity or corner case, or too large number, raising bacteria, time/space compexity 
\section{Motivating Examples}
\label{sec:motivation}

Effective feedback generation for competition-level code requires the ability to apply complex changes to incorrect submissions.
This includes modifying programs’ control flow and making major statement-level alterations.
Along with the ability to perform complex modifications, high repair quality is another priority for \app:  it returns the smallest repairs that it can find.
To illustrate the repair process that \app follows, we use a number of real submissions for Problem 579A from Codeforces as examples.
The prompt for the problem appears in the discussion of Pattern~\ref{para:bacteria} in Section~\ref{para:bacteria}.
%We use the following motivating examples of real-world submissions on Problem 579A from Codeforces to show how \app generates such repairs. The prompt of the question is shown in Pattern~\ref{para:bacteria} in Section~\ref{para:bacteria}.

\begin{figure}[t]
    \centering
    \begin{subfigure}{.35\columnwidth}
    \lstset{escapeinside={<@}{@>}}
	\begin{lstlisting}
<@\textcolor{red}{if}@> ((x/2)!=0)
{
  if ((x%2)==1)
    c++;
  x = x/2;
}
printf("%lld",c+1);
\end{lstlisting}
\caption{Incorrect Program}
\label{subfig:motiv1_incorrect}
    \end{subfigure}%
        \begin{subfigure}{.35\columnwidth}
        \lstset{escapeinside={<@}{@>}}
	\begin{lstlisting}
<@\textcolor{red}{while}@> ((x/2)!=0)
{
  if ((x%2)==1)
    c++;
  x = x/2;
}
printf("%lld",c+1);
\end{lstlisting}
\caption{\app's Repair}
\label{subfig:motiv1_DebugS}
    \end{subfigure}%
        \begin{subfigure}{.33\columnwidth}
        \lstset{escapeinside={<@}{@>}}
	\begin{lstlisting}
<@\textcolor{red}{while(x)}@>
{
  if ((X%2)==1)
    c++;
  x = x/2;
}
printf("%lld",<@\textcolor{red}{c}@>);
  
\end{lstlisting}
\caption{User's Repair}
\label{subfig:motiv1_groundtruth}
    \end{subfigure}%
    \caption{An example repair involving control flow modification.
    The differences between the programs are highlighted in red.
    The variable \texttt{x} is the input for the program, representing the desired number of bacteria to have in the end.
     The variable \texttt{c} is the output, the number of bacteria that need to be inserted.}
    \label{fig:motiv1}
\end{figure}

\para{Incorrect program needs control flow changes.}
An example of a control flow modification that \app applies appears in Figure~\ref{fig:motiv1}.
The original incorrect submission made by a user for Problem 579A appears in Subfigure~\ref{subfig:motiv1_incorrect}.
The high-level design of the implementation is correct, but the control flow needs correction.
Computing the number of ones in the binary representation of the integer \texttt{x} requires a loop rather than a conditional.
Other users in the database repaired their programs by making a similar control flow change (converting \texttt{if} to \texttt{while}), so \app applies the same repair in Subfigure~\ref{subfig:motiv1_DebugS}.
In this situation, \app generates a high-quality repair that not only passes all of the test cases but also makes minimal changes to the structure of the original incorrect program.
The same user's own fix for the problem appears in Subfigure~\ref{subfig:motiv1_groundtruth}.
If we use the Zhang-Shasha algorithm~\cite{zhangshasha} to measure tree edit distances, the repair generated by \app has a distance of one from the original flawed program, whereas the user's own repair has a distance of six from the original program.

\begin{figure}[!t]
    \centering
    \begin{subfigure}{.33\columnwidth}
    \lstset{escapeinside={<@}{@>}}
	\begin{lstlisting}
while (x>0)
{
  if (<@\textcolor{red}{x\%2==0}@>)
    u++;
  x = x/2;
}
printf("%d", u);
\end{lstlisting}
\caption{Incorrect Program}
\label{subfig:motiv2_incorrect}
    \end{subfigure}%
        \begin{subfigure}{.33\columnwidth}
        \lstset{escapeinside={<@}{@>}}
	\begin{lstlisting}
while (x>0)
{
  if (<@\textcolor{red}{x\%2}@>)
    u++;
  x = x/2;
}
printf("%d", u);
\end{lstlisting}
\caption{\app's Repair}
\label{subfig:motiv2_DebugS}
    \end{subfigure}%
        \begin{subfigure}{.33\columnwidth}
        \lstset{escapeinside={<@}{@>}}
	\begin{lstlisting}

while (<@\textcolor{red}{x!=0}@>)
{
  <@\textcolor{red}{u += x\%2}@>; 
  x /= 2;
}
printf("%d",u);
  
\end{lstlisting}
\caption{User's Repair}
\label{subfig:motiv2_groundtruth}
    \end{subfigure}%
    \caption{A example repair involving a statement-level change.
    The differences between the programs are highlighted in red.
    The variable \texttt{x} is the input to this program, and \texttt{u} is the output.}
    \label{fig:motiv2}
\end{figure}

\para{Incorrect program needs statement-level changes.} In addition to making algorithm-level control flow changes, \app is able to generate repairs that require small statement-level changes.
Figure~\ref{fig:motiv2} shows an example of a statement-level repair.
The control flow in the original submission is correct, but the guard in the \texttt{if} statement contains a numerical error.
The repair that \app produces for the submission appears in Subfigure~\ref{subfig:motiv2_DebugS}.
%by learning and applying the similar code changes from the existing programs in the database.
The Zhang-Shasha algorithm gives the new program generated by \app a tree edit distance of only two from the original program.
Although this repair is not the smallest possible repair, which would be changing \texttt{x\%2==0} to \texttt{x\%2==1}, \app still generates a repair that is closer to the user’s original incorrect submission than the user's own repaired program is.
The user's own correction of the program involves three major changes:  changing the guard in the \texttt{while} statement, removing the \texttt{if} statement inside the while loop, and computing the output variable \texttt{u} differently by adding the remainder of \texttt{x} mod 2 to it in each loop iteration.
\section{System Description}
\label{sec:system}
% probably we should mention in the intro that previous research either compare the correct submissions or incorrect submissions.

\begin{comment}
Driven by the insights from our motivating example, we design and build \app, a tool that can automatically generate real-time repair for students' programming contest assignments. 
Our goal is to use the wisdom from the crowd: learning knowledge from how students fixed their programs and apply this knowledge to repair unseen incorrect programs.
To reach this goal, the first step is to efficiently learn and represent code edits from a different version of students' programs into our intermediate representation - merged programs.
Second, \app selects and matches the unseen program with the top-k syntactically closest merged programs.
Third, \app repairs the unseen incorrect program with minimal syntactic changes.
\end{comment}

\begin{figure*}[t!]
\centering\includegraphics[width=.80\linewidth]{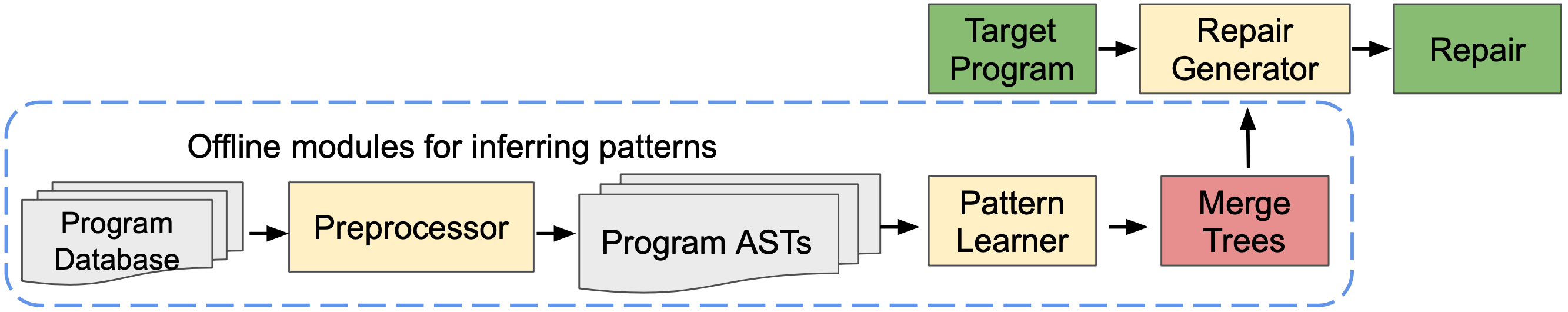}
\caption{\app overview. The green blocks are the input that \app receives from users and the output that it provides for them. The yellow blocks are the key modules of \app.} 
\label{fig:overview}
\end{figure*}

We design and build \app, a tool that can generate repairs for competition-level code automatically by learning the ways that users repair their own programs.
Figure~\ref{fig:overview} gives an overview of \app's architecture. It consists of three main modules:
(1) The preprocessor, described in Section~\ref{subsec:preprocessor}, 
takes the database programs as input,
parses them, and generates
abstract syntax trees for them.
(2) The pattern learner, described in Section~\ref{subsec:repairanalyzer},
uses a new data structure, \textit{merge trees}, to represent the algorithm-level 
and statement-level changes that users in the database apply to their own programs over time.
(3) The repair generator, described in Section~\ref{subsec:repairgenerator},
applies program transformation 
patterns to the incorrect target program to generate repair candidates, 
and it also validates the candidates with the provided test suite for the problem.

\subsection{Preprocessor}
\label{subsec:preprocessor}

The preprocessor parses all of the programs in the database into ASTs offline
for later use in repair pattern learning.
We use the open-source complete C99 parser \textsc{pycparser}~\cite{pycparser} for the AST conversion.
The preprocessor groups the program ASTs into pairs of the form $(i, c)$, where $i$ and $c$ are ASTs for an incorrect program and a correct program, respectively, written by the same user.
If a user made multiple incorrect or correct submissions, the preprocessor makes program pairs for all of the possible combinations.
%The preprocessor selects all of the program pairs: (incorrect, correct) in the database by the same contestants for later examining the differences between these two ASTs.
It also discards programs that have syntax errors in this stage.

\subsection{Pattern Learner}
\label{subsec:repairanalyzer}
In the second module, \app takes the program pairs from the preprocessor as inputs, and it produces a collection of program transformation patterns based on the changes between the incorrect and correct programs.
The program transformation patterns fall into three categories of AST changes:  additions, deletions, and mutations.
\app uses a new data structure, a \textit{merge tree}, to represent the AST changes that occur between the incorrect and correct versions of a program. 

\para{Merge trees.} A merge tree encodes the differences between two abstract syntax trees.
The main structure of a merge tree resembles the unchanged parts of the two ASTs being compared, but the merge tree also includes special nodes that represent additions, mutations, and deletions of sub-trees.
An important characteristic of merge trees is their generality: they can match a variety of patterns in ordinary ASTs rather than only a single pattern.
For example, if the incorrect version of a program contains a statement $c_1$ where the correct version contains a different statement $c_2$, the merge tree for the transformation can apply to ASTs that contain $c_1$, $c_2$, $(c_1;c_2)$, $(c_2;c_1)$, or neither statement, as long as the surrounding parts of the AST bear a sufficiently close resemblance to the merge tree.
In contrast, a simpler encoding of the transformation~\cite{Refazer} would only apply to ASTs that contain $c_1$.
The merge tree's ability to be applied to any program that contains a combination of the two statements makes it cover a much larger range of programs than a simpler encoding does.

\para{Computing program differences.}
The standard approach for identifying differences between two programs is to apply the Zhang-Shasha algorithm directly~\cite{zhangshasha} to compute the edits needed to convert one AST into the other.
Multiple state-of-the-art intro-level feedback generation tools follow this approach~\cite{Refazer,SAR}.
However, the Zhang-Shasha algorithm on its own is not suitable for computing program differences in the domain of competition-level code.
First, the Zhang-Shasha algorithm runs in $\mathcal{O}(m^2n^2)$ time, where $m$ and $n$ are the numbers of nodes in the two input trees.
This imposes a sizable overhead on a tool's operation, hampering its scalability.
%The overhead becomes a bottleneck in computing the difference between two competitive-level code, since the difference between the incorrect and correct versions of a program can be large.
During the development of \app, we found that applying the Zhang-Shasha algorithm to the full ASTs of users' submissions increased the running time of the tool significantly.

The second and more important reason for not using the Zhang-Shasha algorithm on full program ASTs is that the algorithm treats every node in a tree as having equal weight.
The Zhang-Shasha algorithm is a general-purpose algorithm for trees of any kind, not just program ASTs, so it pays no attention to the semantic significance of the edits it uses for measuring distance.
In the case of program ASTs, not all nodes deserve equal weight: some node modifications are more significant than others.  
For example, changing an \texttt{if} node to a \texttt{while} node generally qualifies as a major change because it alters the control flow of a program.  A method for measuring the edit distance between two programs should count such a control flow change as having a higher impact than changing an \texttt{x=1} statement to \texttt{x=0}.
When we compute tree edit distances, we assign a higher cost to control flow edits than to other changes.

%second zss did not associate program edit changes to the control flow
%tomorrow to do
%introduce we first do the cf matching

% our solution is to use control flow compression

%our solution for this is to associate each edit within the control flow node

% need an algo here

\begin{algorithm}
\begin{flushleft}
        \textbf{Input: }$\textit{$P_i$}$ : User's incorrect program submission (AST). \\
        \textbf{Input: }$\textit{$P_c$}$ : User's correct program submission (AST). \\
        \textbf{Output: }$\textit{patternPool}$ : A set of program transformation patterns that reflect the changes that users made in repairing their own programs. \\
\end{flushleft}
\begin{algorithmic}[1]

\Procedure{learnTransformation}{$P_i, P_c$}
	\State \textit{patternPool} = []
	\State \textit{alignedCF}, \textit{unmatchedCF} := ControlFlowAlign($P_i, P_c$)
	\For {$(T_i, T_c)$ \textbf{in} \textit{alignedCF}}
	    %\State $patternPool$ += learnTransformation($T_i, T_c$)
        \State $flatAST_i$, $flatAST_c$ := Flatten($T_i, T_c$)
        \State \textit{edits} := Zhang-Shasha($flatAST_i$, $flatAST_c$)
        \State \textit{mergeTree} += merge(\textit{edits}, $flatAST_i$, $flatAST_c$)
    \EndFor
    \For {$(deletedCF_i)$ \textbf{in} $unmatchedCF$}
        \State $mergeTree$ := augment($deletedCF_i, mergeTree$)
    \EndFor
    \State \textit{patternPool} += \textit{mergeTree}
    
\noindent\text{ }\texttt{ }\texttt{ }\Return \textit{patternPool}
\EndProcedure

\end{algorithmic}
\caption{Learning Program Transformations}
\label{algo:merge}
\end{algorithm}

For \app, we designed a custom algorithm, shown as Algorithm~\ref{algo:merge}, that computes the program transformation patterns between the incorrect and correct versions of a program. 
To detect algorithm-level changes between the two versions, \app uses top-down control flow alignment to capture changes in the control flow.
The nodes that count as control flow nodes for our purposes are \texttt{if}, \texttt{while}, and \texttt{for} statements as well as function calls.
Two control flow nodes align if they have the same type (i.e. they are both \texttt{if}, both \texttt{while}, both \texttt{for}, or both function calls) and they satisfy some extra type-specific conditions for alignment.
Two \texttt{if} statements need to have matching guards or matching true and false branches.
Two \texttt{while} or \texttt{for} statements need to have matching guards or matching bodies.
Two function calls need to have all of their arguments match.
For two sub-expressions to match in any of these cases, they need to be mostly the same.
Variable names and function names are not required to be the same, but the values of variables and the numbers and types of parameters should be.
%their conditions or the statements inside the body of the branch/loop/function are the same\footnote{Variable names and function names are not required to be the same, but the values of variables and the numbers and types of parameters should be.}.
%Since different programs have different variable names, we do not consider variable name differences in alignment.
\app identifies all of the pairs of control flow nodes that align with each other and merges their sub-trees.
%associate changes within the control flow node
%compressed all the children 
%make sure the inner changes not edit on outer nodes
%context sensitive
%changes are associate under the first level of control flow node

A key detail that makes merge tree generation scalable is the fact that the merge tree for a control flow node does not cover changes in the sub-trees of the two versions' control flow nodes.
We leave the sub-trees for other merge trees to cover.
\app handles program changes within the current control flow node by flattening all of the control flow nodes inside its sub-trees and treating the interior as an empty node.
To simplify the process of computing edits, we run the Zhang-Shasha algorithm only on pairs of these flattened sub-trees rather than on the full original ASTs of the incorrect and correct programs.
If $m$ and $n$ are the numbers of nodes of all types in the two input trees and $p$ and $q$ are the numbers of control flow nodes in the two input trees,
then this flattening reduces the time complexity of merge tree generation from $\mathcal{O}(m^2n^2)$ to $\mathcal{O}(\frac{m^2n^2}{p^2q^2})$.

Control flow nodes in the incorrect program with no matches in the correct program are regarded as deletions.
Correspondingly, control flow nodes in the correct program with no matches in the incorrect program are regarded as insertions.
\app uses special nodes as markers in its merge trees to represent these deletions and insertions.
%When there is such similar incorrect control flow detected in the target incorrect program, \app can detect them.
At the end of pattern learning, \app returns a set of all the merge trees it generated to use as program transformation patterns.

\subsection{Repair Generator}
\label{subsec:repairgenerator}

The repair generator takes the incorrect target program and the set of merge trees as input, and it returns a repaired version of the target program.
%The repaired AST is also highlighted in the newly repaired program.
The repair generator's algorithm consists of three main steps.
First, it converts the incorrect target program into an AST just as the preprocessor described in Section~\ref{subsec:preprocessor} does for the database programs.
Second, the repair generator identifies merge trees that match the target program and produces candidate repaired programs by applying transformations based on the merge trees that match the target program.
During this step, variable usage analysis helps with the removal of spurious candidate programs.
%A sound variable alignment analysis is introduced here to remove spurious candidate programs. 
Third, the repair generator validates the candidate programs with the pre-defined test suite for the problem.

\para{Matching the target program with merge trees.}
Merge trees represent the changes that programmers made to their own programs in the database.
The goal of the matching process is to apply similar program edits to repair the target program.
Intuitively, the repair generator takes advantage of the similarities between the incorrect target program and the merge trees to generate repairs.

To start, the repair generator analyzes the target program's AST and identifies several sub-trees that have a control flow node as their root. 
Such a sub-tree matches a merge tree if all of the nodes and edges of the sub-tree are contained within the merge tree.
The names of variables and functions are irrelevant for matching, but the values of variables and the types and parameters of functions matter.
If the repair generator finds a match between the target program and a merge tree, it can modify the target program by performing a repair based on the merge tree.
%This means a similar error made by existing, real-world users is detected in the target incorrect program.
%And the target incorrect program does not contain other nodes that are not contained in the merge tree.
The fact that matching only requires the merge tree to contain the sub-tree and not the other way around helps the repair generator to find ways to fix incorrect submissions in situations where the errors in the submission have no exact matches in the database.
\app generates a modified version of the target program by replacing the matched sub-tree with a new sub-tree based on the merge tree's transformation.
%with the correct tree in the merge tree. 

Replacing a sub-tree using the merge tree's transformation might introduce usages of undefined variables. 
To account for this, the repair generator tries conservatively fitting different combinations of defined variable names onto the undefined variables that are inserted. 
Then it performs variable usage analysis on modified versions of the target program to remove candidates that are invalid simply because of their variable usage.
%since variable names may not be consistent across the sources of all the applied transformations.
The repair generator discards candidate programs that still contain undefined variables after variable alignment or define variables without using them.
This filtering reduces the number of candidate programs to be validated with the test suite, improving the performance of the repair generator.

\para{Validation.} The repair generator validates candidate programs simply by running the provided test suite on them.
As soon as the repair generator finds a candidate program that passes all of the tests, it returns that candidate program as output.

%We rely on a number of heuristics for the process of generating and validating candidate programs.
Because small repairs are more beneficial for users, we prioritize candidates with small transformations over candidates with large ones for validation.
The repair generator starts by applying only one merge tree to the target program at a time to generate candidates.
If we fail to find a valid candidate program after trying every option among the individual merge trees, the repair generator begins creating candidate programs from combinations of multiple transformations.
The repair generator continues trying progressively larger repairs until it hits a timeout or the number of test suite runs reaches a preset limit.
For our evaluation, we do not impose a time limit, but we set a limit of 1,000 on the number of candidate programs to validate with the test suite.

\section{Evaluation}
\label{sec:eval}

We answer the following questions with our evaluation of \app:

\begin{itemize}
\item How effectively can \app repair incorrect submissions for competition-level problems?
\item How high is the quality of \app's feedback?  More specifically, how closely do the repaired programs that \app generates resemble users' original programs?
%\item How helpful is the feedback that \app generates for participants in programming competitions?

\end{itemize}

\setlength{\tabcolsep}{.8em}
\def\arraystretch{1.2}
\begin{table*}[tb]

\centering

\caption{Six Representative Competition-Level Programming Problems.}
\label{table:selected_programs}
\begin{footnotesize}
\resizebox{2.1\columnwidth}{!}{
\begin{tabular}{|c|c|c|c|c|}
\hline
\begin{tabular}[c]{@{}c@{}}Problem \\ ID\end{tabular} & {\color[HTML]{333333} \begin{tabular}[c]{@{}c@{}}Difficulty \\ Level\end{tabular}} & Codeforces Tag           & Challenges                                                          & Problem Description                                                                                                                                                         \\ \hline
1312A                                                 &                                                                                    & Number Theory & Description                                                        & \begin{tabular}[c]{@{}c@{}}Given two integers $n$ and $m$, determine whether a convex regular polygon with $m$ sides can be \\ inscribed in a convex regular polygon with $n$ sides such that their centers and vertices coincide\end{tabular} \\ \cline{1-1} \cline{3-5} 
1519B                                                 & \multirow{-2}{*}{Easy}                                                             & Math          & Description                                                        & \begin{tabular}[c]{@{}c@{}}Given an $n$-by-$m$ grid, with different costs for moving in different directions, \\ check whether it is possible to reach cell $(n,m)$ with exactly cost $k$\end{tabular}           \\ \hline
1238A                                                  &                                                                                    & Number Theory      & \begin{tabular}[c]{@{}c@{}}Algo Design \& \\ Implementation\end{tabular}     & \begin{tabular}[c]{@{}c@{}}Given two integers $x$ and $y$, determine whether there is a prime integer $p$ \\ such that subtracting $p$ from $x$ any number of times makes $x$ equal to $y$\end{tabular}                                  \\ \cline{1-1} \cline{3-5} 
1295A                                                & \multirow{-2}{*}{Medium}                                                           & Greedy      & \begin{tabular}[c]{@{}c@{}}Description \& \\ Algo Design\end{tabular}     & \begin{tabular}[c]{@{}c@{}}Find the largest integer that can be shown on a seven-segment (alarm clock) \\ display that requires no more than $n$ segments to be turned on in total\end{tabular}                                                                                     \\ \hline
579A                                                   &                                                                                    & Bit Mask      & \begin{tabular}[c]{@{}c@{}}Algo Design \& \\ Implementation\end{tabular}     &
\begin{tabular}[c]{@{}c@{}}Find the minimum number of bacteria that need to be placed into a box over \\ some number of days in order to have $x$ bacteria in the box at some moment  \end{tabular}                                                                                                   \\ \cline{1-1} \cline{3-5} 
1199B                                                 & \multirow{-2}{*}{Hard}                                                           & Geometry      & \begin{tabular}[c]{@{}c@{}}Algo Design \& \\ Implementation\end{tabular}     & \begin{tabular}[c]{@{}c@{}} Find the depth of a body of water given the distance that a vertical line \\ segment extending from the bottom can tilt before being submerged\end{tabular}                                                                                                                                       \\ \hline

\end{tabular}
}
\end{footnotesize}
\end{table*}

\subsection{Implementation and Experimental Setup}
Our implementation of \app uses a mix of Python and open-source software libraries.
%We have built a \app prototype using a mix of Python and open-source software libraries.
As it is now, \app operates on C programs.
We rely on \textsc{pycparser}~\cite{pycparser}, a complete C99 parser, to convert C programs into abstract syntax trees.
Also, we use the Zhang-Shasha algorithm~\cite{zss} to compute tree edit distances.
%All of the experiments in this section were conducted on a MacBook Pro equipped with an Intel i7 CPU, 16GB of memory, and a PCIe-based 512 GB SSD. 
% Detailed evaluation on how \app repair five programs

\setlength{\tabcolsep}{.8em}
\def\arraystretch{1.22}
\begin{table*}[tb]

\centering

\caption{Statistics for the six selected competition-level programming problems.
The categorizations for submissions here come from Codeforces. AC: Accepted, WA: Wrong Answer, CE: Compile-Time Error, RE: Runtime Error, TLE: Time Limit Exceeded, MLE: Memory Limit Exceeded, OT: Other.}
\label{table:comparision_stat}
\begin{footnotesize}
\resizebox{1.7\columnwidth}{!}{
\begin{tabular}{|c|c|c|c|c|c|c|c|c|c|}
\hline
Problem ID & \# Submissions & \# AC & \# WA  &  \# CE  & \# RE & \# TLE & \#  MLE & \# OT & Average LOC  \\ \hline
 1312A &   1160    &   494         &  327   & 301     & 15   &  19   &  3   &   1  & 21.1     \\ \hline
1519B                              &  724 &   349       & 211    &  137   &  12  &   14  &  1   &  0       & 25.1    \\ \hline
1238A                                &  1345 &   303       &  520   & 358    &  39  & 121    &  0   & 4       & 27.3      \\ \hline
1295A                                &  1024 & 251         &     349 &  368   &  13  & 40     &    3 & 0         & 33.3     \\ \hline
579A                        & 1889       & 780        & 758 & 288  & 23  & 36  & 1   & 3       & 19.8   \\ \hline
1199B                     & 2045         &    413        &   1339  &  269   & 18   & 1    & 0     &    5     & 11.9    \\  %\hhline{|=|=|=|=|=|=|=|=|=|=|}
%\hhline{|-|-|-|-|-|-|-|-|-|-|}
%\hhline{|-|-|-|-|-|-|-|-|-|-|}
\hline\hline
Total & 8187 & 2590 &	3504 &	1721 &	120 &	231 &	8 &	13 &  21.4 \\ \hline
\end{tabular}
}
\end{footnotesize}
\end{table*}

\para{Benchmark Setup.} For our evaluation suite, we use six problems from Codeforces,\footnote{https://codeforces.com} the world's largest online platform for competitive programming.
Codeforces assigns difficulty scores to its problems, and we group the problems into three categories based on their difficulty scores.
We categorize problems with a difficulty score of 800 or less as \emph{easy}, problems with a score between 800 and 1000 as \emph{medium}, and problems with a score of at least 1000 as \emph{hard}.
Each of the six problems that we selected received more than 700 submissions written in C.
(Multiple submissions made by the same user count as distinct.)
For each submission, we collect not only the text of the program but also its execution result, running time, and memory usage.\footnote{The founders and owners of Codeforces gave us permission to collect data from the six problems.}
Additionally, we have access to the official test suite used by Codeforces for each of the problems.
Table \ref{table:selected_programs} names the six problems and the specific challenges that each problem presents.

Table \ref{table:comparision_stat} provides a breakdown of the six selected competition-level programming problems.
We can see from analyzing the execution results for the database programs that 25.6\% (2093/8187) of the submissions were rejected because of errors rather than incorrect outputs.
Among those, 4.5\% (370/8187) of the programs were classified as incorrect because of runtime errors, time limit violations, or memory limit violations.
Furthermore, of the 5597 incorrect submissions, 2921 (\abandonrate) come from programmers who never made a correct submission.
%In this, 1437 of them have compiler error, we collected the rest of submissions into our evaluation target. 
%are incorrect due to the errors found in the runtime, time limit exceeded or memory limit exceeded, which is the result of rigorous checks on execution time and memory consumption in competitive-level code.

%need to include the statistic of the benchmark to highlight this problem, e.g., many users ended up being incorrect and abandon the problem.
%maybe in the intro or the second paragraph to make sure it fly

\app aims to provide effective feedback for programmers as they practice on competition-level problems.
To assess whether \app meets this goal, we split the users into two groups for each programming problem:
\begin{itemize}
\item \textbf{Group One.} Some users made one or more incorrect submissions but never managed to produce a correct submission.
%For this group, we check whether \app can create a repaired, correct version of a user's final incorrect submission at all.
Generating repairs for these users' programs is generally a challenging task:  since the users never managed to repair their own programs, the submissions may contain major errors.
\item \textbf{Group Two.} Other users made one or more incorrect submissions initially but then managed to produce a correct submission afterward on their own.
These users' incorrect submissions are easier to handle in general:  the fact that the users found a solution eventually means that they were likely close to a right answer with their earlier incorrect submissions.
%For these users, we check whether \app can produce a repaired program that is closer to the user's final incorrect submission than the same user's correct submission is in terms of program editing distance.
%(We view the user's correct submission as the ground truth for the problem of repairing the incorrect submission.)
\end{itemize}

For each problem, we perform some cleaning of the database before we use it as a training set.
To clean the database, we discard all programs with syntax errors and all submissions from users who solved the problem on their first attempt.
Next, we label each user with a distinct anonymous identifier.
After that, we allocate 80\% of the users for the training set and 20\% for the evaluation set.
Some existing feedback generators use the chronologically earliest 80\% of submissions as the training set and the remaining 20\% as the evaluation set~\cite{SKP}, but we divide the users at random instead to avoid skewing the results.
The setup of Codeforces makes problems easier for programmers who submit their programs later:  on Codeforces, users practicing on a specific problem can view every other user's submission history for the same problem.
Consequently, there is a risk that programmers who submitted later fixed the mistakes in their code by copying someone else's correct submission rather than by finding a solution on their own.
Grouping the users randomly rather than chronologically allows us to distribute the users who copied other users' submissions more fairly between the training set and the evaluation set, if there are any such users.

%For each problem, we disregarded the programs with syntax errors in our training set. 
%We used distinct, anonymous identifiers to mark each user.
%In Codeforces, all the submissions are public to each user.
%Thus, to avoid the possible threat that the later programmers pasted the chronologically earlier correct program as their own solution, we shuffled the users (with their submissions) alphabetically using anonymous identifiers.
%After that, we collected the users who passed this problem within the first attempt.
%Then we split the remaining users into two categories.
%In Group one, we further randomly selected the first 80\% users as the train and the rest of 20\% users as the testing to \app.

\setlength{\tabcolsep}{.8em}
\def\arraystretch{1.2}
\begin{table*}[tb]

\centering

\caption{Evaluation of \app on incorrect programs abandoned by their authors (Group One). The second column shows the number of incorrect-correct program pairs in the training set for each problem, not the number of individual programs.
The penultimate column shows the average dissimilarity of the target programs that had a successful repair by \app. The last column shows the average dissimilarity of the target programs that \app failed to generate a successful repair.
%The closeness is defined by the tree editing distance between this target program to the closest correct program in the database.
}
\label{table:comparision_result1}
\begin{footnotesize}
\resizebox{2.1\columnwidth}{!}{
\begin{tabular}{|c|c|c|c|c|c|c|c|c|}
\hline
{\color[HTML]{000000} \begin{tabular}[c]{@{}c@{}}Problem \\ ID\end{tabular}} & \begin{tabular}[c]{@{}c@{}} \# Pairs in \\ Training Set \end{tabular} & \begin{tabular}[c]{@{}c@{}} \# Programs \\ in Test Set\end{tabular} & \begin{tabular}[c]{@{}c@{}} \# Programs \\ Repaired\end{tabular} & \begin{tabular}[c]{@{}c@{}}Accuracy\\ (Repair Rate)\end{tabular} & \begin{tabular}[c]{@{}c@{}}Avg. Relative\\ Repair Size\end{tabular} & \begin{tabular}[c]{@{}c@{}} Average \\ Dissimilarity \end{tabular} & \begin{tabular}[c]{@{}c@{}}Avg. Dissimilarity\\ for Successes \end{tabular} & \begin{tabular}[c]{@{}c@{}}Avg. Dissimilarity \\ for Failures\end{tabular} \\ \hline
{\color[HTML]{000000} 1312A}       & 475                                                                                  & 90         & 51                                                         & 56.7\%                                                           & 0.24                                                                & 0.26                                                                                       & 0.16                                                                                        & 0.39                                                                                   \\ \hline
1519B                              & 203                                                                                  & 31         & 12                                                         & 38.7\%                                                           & 0.38                                                                & 0.43                                                                                       & 0.27                                                                                        & 0.53                                                                                   \\ \hline
1238A                              & 1304                                                                                 & 316        & 119                                                        & 37.7\%                                                           & 0.26                                                                & 0.40                                                                                       & 0.27                                                                                        & 0.48                                                                                   \\ \hline
1295A                              & 277                                                                                  & 127        & 38                                                         & 29.9\%                                                           & 0.86                                                                & 0.56                                                                                       & 0.55                                                                                        & 0.57                                                                                   \\ \hline
579A                               & 1654                                                                                 & 362        & 98                                                         & 27.1\%                                                           & 0.73                                                                & 0.44                                                                                       & 0.41                                                                                        & 0.45                                                                                   \\ \hline
1199B                              & 4027                                                                                 & 558        & 82                                                         & 14.7\%                                                           & 0.15                                                                & 0.19                                                                                       & 0.06                                                                                        & 0.21                                                                                   \\ \hline
\end{tabular}
}
\end{footnotesize}
\end{table*}

% line of code includes the header files but for (all the files in both training and testing set)
% 
\setlength{\tabcolsep}{.8em}
\def\arraystretch{1.2}
\begin{table*}[tb]

% may need precision and recall
\centering

\caption{Evaluation of \app on incorrect programs later repaired by their authors (Group Two).
The training set for each problem is the same as it is in Table \ref{table:comparision_result1}.
%A repair generated by \app is defined as a high-quality repair if it is syntactically closer to original target incorrect program than user's later own repair (ground truth).
}
\label{table:comparision_result2}
\begin{footnotesize}
\resizebox{2.1\columnwidth}{!}{
\begin{tabular}{|c|c|c|c|c|c|c|c|c|}
\hline
{\color[HTML]{000000} \begin{tabular}[c]{@{}c@{}}Problem \\ ID\end{tabular}} & \begin{tabular}[c]{@{}c@{}} \# Programs \\ in Test Set\end{tabular} & \begin{tabular}[c]{@{}c@{}} \# Programs \\ Repaired\end{tabular} & \begin{tabular}[c]{@{}c@{}}Accuracy\\ (Repair Rate)\end{tabular} & \begin{tabular}[c]{@{}c@{}}  High-Quality\\ Repairs\end{tabular} & \begin{tabular}[c]{@{}c@{}}Avg. Relative\\ Repair Size\end{tabular} & \begin{tabular}[c]{@{}c@{}}Average \\ Dissimilarity\end{tabular} & \begin{tabular}[c]{@{}c@{}}Avg. Dissimilarity\\ for Successes \end{tabular} & \begin{tabular}[c]{@{}c@{}}Avg. Dissimilarity\\ for Failures\end{tabular} \\ \hline
{\color[HTML]{000000} 1312A}                                                                                                       & 62                                                           & 44                                                         & 71.0\%                                                           & 52.3\%                                                                         & 0.19                                                                & 0.17                                                        & 0.14                                                                          & 0.21                                                                     \\ \hline
1519B                                                                                                                              & 71                                                           & 55                                                         & 77.5\%                                                           & 7.3\%                                                                          & 0.33                                                                & 0.19                                                        & 0.12                                                                          & 0.40                                                                     \\ \hline
1238A                                                                                                                                  &   55                                                           &        34                                                    &     61.8\%                                                             &  58.8\%                                                                       &          0.29                                                                  &           0.34                                                  &          0.24                                                                     &               0.50                                                           \\ \hline
1295A                                                                                                                                  &             53                                                 &        22                                                    &      41.5\%                                                            &          13.6\%                                                                      &      0.54                                                               &             0.38                                                &           0.37                                                                    &                0.39                                                          \\ \hline
579A                                                                                                                             & 107                                                          & 25                                                         & 27.1\%                                                           & 52.0\%                                                                         & 0.53                                                                & 0.40                                                        & 0.38                                                                          & 0.41                                                                     \\ \hline
1199B                                                                                                                           & 176                                                          & 37                                                         & 21.0\%                                                           & 59.5\%                                                                         & 0.21                                                                & 0.15                                                        & 0.08                                                                          & 0.17                                                                     \\ \hline
\end{tabular}
}
\end{footnotesize}
\end{table*}

\subsection{Results}
\label{subsec:analysis}

\para{Group One.} Table \ref{table:comparision_result1} shows \app's results for incorrect programs abandoned by their authors. 
To this group, \app has an overall fix rate of \pureincorrectaccuracy across the six problems. Since the programs' authors never addressed their mistakes fully on their own, they would benefit from receiving repaired versions of their programs as feedback.
%We cannot measure the quality of our repairs as we do for Group Two because we have no correct program versions written by the same users.
To measure the quality of our repairs for Group One, we introduce a new metric:  the \textit{dissimilarity} between each target program and the correct programs in the database.
%Instead, we measure repair quality with a new metric:  the \textit{closeness} to measure how close this incorrect program is to any of the correct programs.
To measure dissimilarity, we compute the minimum tree edit distance between a target program and any of the correct programs in the database using the Zhang-Shasha algorithm, and then we divide this distance by the size of the target program. 
This dissimilarity metric quantifies the difficulty of repairing each program:  if a program is not syntactically close to any correct program in the database, generating a repair for it is difficult, and any repairs found are likely to be large.

\app generates high-quality repairs for Group One according to our standard.
We define the \textit{relative repair size} for a problem as the tree edit distance between a repaired program and its target program divided by the size of the target program.
For three of the six problems, \app has an average relative repair size smaller than the average dissimilarity between the target programs and the correct programs.
The high dissimilarity values for Group One make the higher average relative repair sizes for the other three problems understandable.
%Another important finding is that \app fails to generate repairs when the target program is largely incorrect.
Another important finding is that the average dissimilarity across all six problems is 0.44, so a typical target program needs a large portion of its code to be changed to become identical to any of the correct database programs.

\para{Group Two.}
Table \ref{table:comparision_result2} shows \app's results for incorrect programs later fixed by their authors.
\app has an overall fix rate of \hasrepairaccuracy across the six problems, which is better than the result for Group One.
Since we have the authors' own repairs for the programs in Group Two, we use the authors' repairs as the ground truth for assessing the quality of \app's feedback.
%All of the programs used in this experiment have the user's original repairs.
A repair generated by \app counts as a \textit{high-quality repair} if the tree edit distance between it and the target program is smaller than the tree edit distance between the user's own repair and the target program.
For four out of the six problems, more than 50\% of the repairs \app generates are closer to the target program than the ground truth is, so \app does in fact generate high-quality repairs for programs in Group Two.

% Detailed evaluation on how Clara repairs these five programs

%we need to emphasize that our fix is based on the user changed part, this is something that Clara cannot do!

%\subsection{Comparison with the state-of-the-art: Clara}

\subsection{Threats to Validity}
\para{Internal validity.}
\app validates candidate programs by running the test suite provided by Codeforces on them.
Passing every test within the imposed memory and time limits is not a perfect guarantee of the correctness of a program but only a highly likely indicator of its correctness.
A perfect guarantee would require formal verification, which we do not perform.
To our knowledge, this limitation is common to all existing data-driven feedback generation techniques~\cite{Refazer,clara,SAR,Refactory,FAPR}.

\para{External validity.}
Currently, \app only supports feedback generation for C programs. However, the general principles behind the design of \app are applicable to programs written in any language.
Our method for handling control flow nodes does assume C-like syntax, but nothing else about the underlying algorithm is tailored specifically for C.

\section{Discussion}
\label{sec:discussion}

\para{State-of-the-art tools.}
Recent data-driven approaches for feedback generation utilize the wisdom of the crowd by selecting \emph{donor programs} from their databases~\cite{clara,SAR,Refactory,Refazer}.
A donor program is a program that bears a close resemblance to the program to be repaired.
Tools that use donors repair their target programs by analyzing the differences between a target program and its donors.
%Different tools rely on different definitions of program similarity, so the resemblance between the donor programs and the target program can be syntactic, semantic, or both.

State-of-the-art feedback generators choose their donor programs from a database of programs that are either all correct or all incorrect.
Both options have limitations.
Tools that draw their donors from databases of correct programs~\cite{clara,SAR,Refactory} operate under the faulty assumption that the target program differs from the correct database programs
%, either syntactically or semantically,
only because of the presence of errors in the program.
%If the target program differs significantly from the correct database programs and is functionally incorrect, the tools will fail to identify matches for the program and consequently will fail to generate repairs for it.
Tools that draw their donors from databases of incorrect programs~\cite{Refazer} suffer from low success rates because
%generate repairs successfully only when the errors in the donor programs bear a sufficiently close resemblance to the errors in the target program.
the mistakes in the donor programs are unlikely to coincide with the mistakes in the target program.
%, so these tools suffer from a low success rate.
%Additionally, the overall structure of the target program may differ from the overall structure of the donor programs.
%Tools with only incorrect programs in their databases generate repairs successfully only when the errors in the donor programs bear a sufficiently close resemblance to the errors in the target program.

\app takes a different approach.  Its database includes both correct programs and incorrect programs, and it draws information from both sides of the database to produce merge trees that function like donor programs do for other tools.
Merge trees offer a unique advantage over donor programs:  they allow \app to generate high-quality repairs in a way that mimics the debugging procedure that human programmers follow.
The structure of a merge tree represents the changes that a user makes between the different versions of a program.  \app can learn to imitate the user's behavior by observing the differences between an early incorrect version of the program and the final correct version of the program.

\para{Direct comparison.}
A number of factors prevent us from performing a direct comparison between \app and any existing feedback generator.
Language compatibility is a major issue:  the only publicly available existing feedback generator that operates on C programs is Clara~\cite{clara}.
We cannot evaluate Clara on the same problems from Codeforces that we used for \app because Clara does not support the full range of C's syntax.
Expressions such as \texttt{while(t\textminus\textminus)} or \texttt{a[i]++}, where \texttt{a} is an array of integers, cause Clara to return an error.
%Instead of attempting to generate feedback, Clara returns an error if it encounters any such problematic expression.
Also, Clara takes only a single C function at a time as input rather than an entire program.
Reformulating every target program as a single function to circumvent this problem is not an option:  function definitions within the original program would become nested function definitions, which Clara does not support.
Lastly, the most important reason for not performing a direct comparison against Clara is that Clara operates under a different definition of correctness than \app does.
Clara regards all functionally correct programs as valid, even if they use too much time or memory.
A comparison of the success rates of \app and Clara on the same set of programs would not be very informative because the two tools' reported success rates mean different things.

\section{Related Work}
\para{Competitive programming.}
Researchers have devoted an increasing amount of attention to competitive programming in recent years because of its growing impact on programming training and education~\cite{bloomfield2016programming,CompetitiveProgrammingGuide}.
Laaksonen~\cite{CompetitiveProgrammingGuide} provides a systematic guide to algorithm design strategy in competitive programming.
Puri {\em et al.}~\cite{Puri} provide a large database of thousands of competitive programming problems along with millions of sampled solutions for the problems.
The first tool to generate solutions for programming problems with program synthesis comes from Zavershynskyi {\em et al.}~\cite{NAPS}.
Unfortunately, the tool's utility is limited significantly by the fact that it generates solutions only in a custom-made intermediate programming language.
Hendrycks {\em et al.}~\cite{APPS} are the first to use large language models to generate solutions for competitive programming problems directly in Python.
However, their approach produces solutions successfully less than 10\% of the time.
AlphaCode~\cite{alphacode} is a significant improvement over the state of the art~\cite{NAPS,APPS,codex}.
AlphaCode produces programs based on natural-language descriptions that it receives as input.  In contests with more than 5,000 participants, AlphaCode places among the top 54.3\% of participants on average~\cite{alphacode}.
In spite of the advances that researchers have made in the field of competitive programming, no existing tool generates feedback or repairs for incorrect competition-level programs.

\para{Automated feedback generation.}
Automatic feedback generation for programming assignments has been a popular topic in programming education over the last decade~\cite{Autograder,CoderAssist,QLOSE,SKP,Refazer,YiFSE17,wang2018dynamic,clara,SAR,Refactory,SemCluster,FAPR,Cafe}.
The first tools developed for the task~\cite{Autograder,CoderAssist}
rely on manual guidance from users, in the form of either reference solutions~\cite{Autograder,CoderAssist} or an error model~\cite{Autograder} that explicitly defines all of the repairs that the tool can make.
Because of their heavy reliance on input from users, early feedback generation tools do not qualify as fully automatic.

More recent feedback generators do qualify as fully automatic, and they rely on data-driven approaches for the task.  They learn how to generate repairs for programs by analyzing programs written by other users.
Tools such as Clara~\cite{clara}, SARFGEN~\cite{SAR}, Refactory~\cite{Refactory}, FAPR~\cite{FAPR}, and Cafe~\cite{Cafe} use databases of existing correct solutions for a problem to learn how to repair incorrect programs written for the same problem.
Some of the data-driven tools are limited by their heavy dependence on syntactic similarities between the target program and reference solutions from the database.
Two of the tools for imperative languages cannot repair a flawed program unless their database contains a correct program with exactly the same control flow as the flawed program~\cite{clara,SAR}.
Similarly, one of the tools for functional programs requires alignment for function call sites~\cite{Cafe}.
Multiple studies have shown that the assumption that a flawed program will have an exact control-flow match in the database of correct programs is too strong to be reliable~\cite{FixML,Refactory}.
Other feedback generators suffer from different problems, such as the tendency to enlarge programs excessively with their repairs~\cite{clara,Refactory}, the inability to fix errors that require changes to multiple parts of a program~\cite{Refazer}, and the inability to take programs' semantics into consideration~\cite{BIFI,FAPR}.

Furthermore, state-of-the-art feedback generators~\cite{clara,SAR,Cafe} cannot generate the complex repairs that flawed competition-level programs need because the tools' creators designed them with intro-level programming assignments in mind.
No existing tool can repair programs that require an algorithm-level redesign, but merge trees allow \app to handle the task.
The inspiration behind our usage of merge trees comes from algorithms for semi-structured merging~\cite{SvenSemistructured,CavalcantiSemistructured}.
More importantly, no existing feedback generator attempts to make programs more efficient.
%Existing tools view correctness only in terms of input-output correspondence, not a program's time or memory usage.

\para{Automated program repair.}
%\app is also closely related to automated program repair techniques.
Researchers have studied automated program repair techniques extensively for the past sixty years~\cite{APRGoues13,APRGoues19,genesis,prophet}.
Automated program repair techniques fall into three main categories:  heuristic-based techniques~\cite{GenProg,RandomQi,PAR}, semantics-based techniques~\cite{Angelix,MechtaevNNGR18}, and learning-based techniques~\cite{prophet,genesis}.
Heuristic-based approaches use some heuristic, such as genetic programming~\cite{GenProg}, randomization~\cite{RandomQi}, or a predefined fitness function~\cite{PAR}, to guide a search procedure to candidate patches for a program.
Semantics-based techniques~\cite{Angelix,MechtaevNNGR18} combine symbolic execution with SMT solvers to synthesize repairs.
Semantics-based techniques cannot repair competition-level code reliably because of the limitations of their internal design.
Programming competitions make heavy use of floating-point numbers for geometry problems and lists for string operation problems, both of which are difficult for SMT solvers to handle effectively.
Learning-based techniques~\cite{prophet,genesis} learn code repair patterns from prior patches. 

State-of-the-art automated program repair techniques work best when used to handle a small number of errors in a large code base that includes millions of lines of code.
Consequently, these techniques are impractical for competition-level code, where errors appear much more frequently relative to the size of users' programs.

Automatic repair for non-functional program properties (i.e. time and memory usage) has received a small amount of attention from researchers previously.
However, unlike \app, prior work in the area has targeted only specific program patterns~\cite{APRGoues13,APRGoues19}, such as unnecessary loop iterations~\cite{Caramel} or repeated computations of the same value~\cite{Cachetor}.
No prior research on the subject has led to the development of a general-purpose tool for improving the efficiency of competition-level code automatically.
\section{Conclusion}
We present \app, a tool that generates feedback automatically for competition-level code.
By observing how other users repair their own programs over time, \app learns how to create repairs for its target programs.
The improvement in quality that \app provides over the standard feedback that programmers receive when practicing on competition-level problems will make online programming platforms that utilize \app more user-friendly.
%Inspired by the fact many users abandoned the problem after several unsuccessful attempts due to insufficient feedback from the competition platform, \app can automatically generate repairs for the users.
\bibliography{repair}


\begin{thebibliography}{46}

%%% the Web address.

\ifx \showCODEN    \undefined \def \showCODEN     #1{\unskip}     \fi
\ifx \showDOI      \undefined \def \showDOI       #1{#1}\fi
\ifx \showISBNx    \undefined \def \showISBNx     #1{\unskip}     \fi
\ifx \showISBNxiii \undefined \def \showISBNxiii  #1{\unskip}     \fi
\ifx \showISSN     \undefined \def \showISSN      #1{\unskip}     \fi
\ifx \showLCCN     \undefined \def \showLCCN      #1{\unskip}     \fi
\ifx \shownote     \undefined \def \shownote      #1{#1}          \fi
\ifx \showarticletitle \undefined \def \showarticletitle #1{#1}   \fi
\ifx \showURL      \undefined \def \showURL       {\relax}        \fi
% invisible to TeX
\providecommand\bibfield[2]{#2}
\providecommand\bibinfo[2]{#2}
\providecommand\natexlab[1]{#1}
\providecommand\showeprint[2][]{arXiv:#2}

\bibitem[\protect\citeauthoryear{??}{odd}{2022}]%
        {oddDivisorStackoverflow}
 \bibinfo{year}{2022}\natexlab{}.
\newblock \bibinfo{title}{An example of repairing a faulty submission that need
  an algorithm-level redesign.}
\newblock
\newblock
\urldef\tempurl%
\url{https://stackoverflow.com/questions/65896295/finding-odd-divisors-with-bit-shifting}
\showURL{%
\tempurl}


\bibitem[\protect\citeauthoryear{??}{hpe}{2022}]%
        {hpecodewars}
 \bibinfo{year}{2022}\natexlab{}.
\newblock \bibinfo{title}{{HPE CODEWARS}}.
\newblock \bibinfo{howpublished}{\url{https://hpecodewars.org/}}.
\newblock


\bibitem[\protect\citeauthoryear{??}{mic}{2022}]%
        {microsoftImagineCup}
 \bibinfo{year}{2022}\natexlab{}.
\newblock \bibinfo{title}{{Microsoft Imagine Cup}}.
\newblock \bibinfo{howpublished}{\url{https://imaginecup.com/}}.
\newblock


\bibitem[\protect\citeauthoryear{??}{pyc}{2022}]%
        {pycparser}
 \bibinfo{year}{2022}\natexlab{}.
\newblock \bibinfo{title}{{pycparser: Complete C99 parser in pure Python}}.
\newblock \bibinfo{howpublished}{\url{https://github.com/eliben/pycparser}}.
\newblock


\bibitem[\protect\citeauthoryear{??}{com}{2022}]%
        {companyContest}
 \bibinfo{year}{2022}\natexlab{}.
\newblock \bibinfo{title}{{The 10 Most Prestigious Programming Contests and
  Coding Challenges}}.
\newblock
  \bibinfo{howpublished}{\url{https://www.mycplus.com/featured-articles/programming-contests-and-challenges/}}.
\newblock


\bibitem[\protect\citeauthoryear{??}{Fac}{2022}]%
        {FacebookHacker}
 \bibinfo{year}{2022}\natexlab{}.
\newblock \bibinfo{title}{{The Facebook Hacker Cup}}.
\newblock
  \bibinfo{howpublished}{\url{https://www.facebook.com/codingcompetitions/hacker-cup/}}.
\newblock


\bibitem[\protect\citeauthoryear{??}{ICP}{2022}]%
        {ICPC}
 \bibinfo{year}{2022}\natexlab{}.
\newblock \bibinfo{title}{{The International Collegiate Programming Contest}}.
\newblock \bibinfo{howpublished}{\url{https://icpc.global/}}.
\newblock


\bibitem[\protect\citeauthoryear{??}{Yan}{2022}]%
        {Yandexalgo}
 \bibinfo{year}{2022}\natexlab{}.
\newblock \bibinfo{title}{{The Yandex Algorithm Cup}}.
\newblock \bibinfo{howpublished}{\url{https://yandex.com/cup/algorithm/}}.
\newblock


\bibitem[\protect\citeauthoryear{Apel, Liebig, Brandl, Lengauer, and
  K\"{a}stner}{Apel et~al\mbox{.}}{2011}]%
        {SvenSemistructured}
\bibfield{author}{\bibinfo{person}{Sven Apel}, \bibinfo{person}{J\"{o}rg
  Liebig}, \bibinfo{person}{Benjamin Brandl}, \bibinfo{person}{Christian
  Lengauer}, {and} \bibinfo{person}{Christian K\"{a}stner}.}
  \bibinfo{year}{2011}\natexlab{}.
\newblock \showarticletitle{Semistructured Merge: Rethinking Merge in Revision
  Control Systems}. In \bibinfo{booktitle}{\emph{Proceedings of the 19th ACM
  SIGSOFT Symposium and the 13th European Conference on Foundations of Software
  Engineering}} (Szeged, Hungary) \emph{(\bibinfo{series}{ESEC/FSE '11})}.
  \bibinfo{publisher}{Association for Computing Machinery},
  \bibinfo{address}{New York, NY, USA}, \bibinfo{pages}{190–200}.
\newblock
\showISBNx{9781450304436}
\urldef\tempurl%
\url{https://doi.org/10.1145/2025113.2025141}
\showDOI{\tempurl}


\bibitem[\protect\citeauthoryear{Bloomfield and Sotomayor}{Bloomfield and
  Sotomayor}{2016}]%
        {bloomfield2016programming}
\bibfield{author}{\bibinfo{person}{Aaron Bloomfield} {and}
  \bibinfo{person}{Borja Sotomayor}.} \bibinfo{year}{2016}\natexlab{}.
\newblock \showarticletitle{A programming contest strategy guide}. In
  \bibinfo{booktitle}{\emph{Proceedings of the 47th ACM technical symposium on
  computing science education}}. \bibinfo{pages}{609--614}.
\newblock


\bibitem[\protect\citeauthoryear{Boyer and Moore}{Boyer and Moore}{1991}]%
        {Moore91}
\bibfield{author}{\bibinfo{person}{Robert~S. Boyer} {and}
  \bibinfo{person}{J.~Strother Moore}.} \bibinfo{year}{1991}\natexlab{}.
\newblock \showarticletitle{{MJRTY:} {A} Fast Majority Vote Algorithm}. In
  \bibinfo{booktitle}{\emph{Automated Reasoning: Essays in Honor of Woody
  Bledsoe}} \emph{(\bibinfo{series}{Automated Reasoning Series})},
  \bibfield{editor}{\bibinfo{person}{Robert~S. Boyer}} (Ed.).
  \bibinfo{publisher}{Kluwer Academic Publishers}, \bibinfo{pages}{105--118}.
\newblock


\bibitem[\protect\citeauthoryear{Cavalcanti, Borba, and Accioly}{Cavalcanti
  et~al\mbox{.}}{2017}]%
        {CavalcantiSemistructured}
\bibfield{author}{\bibinfo{person}{Guilherme Cavalcanti},
  \bibinfo{person}{Paulo Borba}, {and} \bibinfo{person}{Paola Accioly}.}
  \bibinfo{year}{2017}\natexlab{}.
\newblock \showarticletitle{Evaluating and Improving Semistructured Merge}.
\newblock \bibinfo{journal}{\emph{Proc. ACM Program. Lang.}}
  \bibinfo{volume}{1}, \bibinfo{number}{OOPSLA}, Article
  \bibinfo{articleno}{59} (\bibinfo{date}{oct} \bibinfo{year}{2017}),
  \bibinfo{numpages}{27}~pages.
\newblock
\urldef\tempurl%
\url{https://doi.org/10.1145/3133883}
\showDOI{\tempurl}


\bibitem[\protect\citeauthoryear{Chen, Tworek, Jun, Yuan, Pinto, Kaplan,
  Edwards, Burda, Joseph, Brockman, Ray, Puri, Krueger, Petrov, Khlaaf, Sastry,
  Mishkin, Chan, Gray, Ryder, Pavlov, Power, Kaiser, Bavarian, Winter, Tillet,
  Such, Cummings, Plappert, Chantzis, Barnes, Herbert-Voss, Guss, Nichol,
  Paino, Tezak, Tang, Babuschkin, Balaji, Jain, Saunders, Hesse, Carr, Leike,
  Achiam, Misra, Morikawa, Radford, Knight, Brundage, Murati, Mayer, Welinder,
  McGrew, Amodei, McCandlish, Sutskever, and Zaremba}{Chen
  et~al\mbox{.}}{2021}]%
        {codex}
\bibfield{author}{\bibinfo{person}{Mark Chen}, \bibinfo{person}{Jerry Tworek},
  \bibinfo{person}{Heewoo Jun}, \bibinfo{person}{Qiming Yuan},
  \bibinfo{person}{Henrique Ponde de~Oliveira Pinto}, \bibinfo{person}{Jared
  Kaplan}, \bibinfo{person}{Harri Edwards}, \bibinfo{person}{Yuri Burda},
  \bibinfo{person}{Nicholas Joseph}, \bibinfo{person}{Greg Brockman},
  \bibinfo{person}{Alex Ray}, \bibinfo{person}{Raul Puri},
  \bibinfo{person}{Gretchen Krueger}, \bibinfo{person}{Michael Petrov},
  \bibinfo{person}{Heidy Khlaaf}, \bibinfo{person}{Girish Sastry},
  \bibinfo{person}{Pamela Mishkin}, \bibinfo{person}{Brooke Chan},
  \bibinfo{person}{Scott Gray}, \bibinfo{person}{Nick Ryder},
  \bibinfo{person}{Mikhail Pavlov}, \bibinfo{person}{Alethea Power},
  \bibinfo{person}{Lukasz Kaiser}, \bibinfo{person}{Mohammad Bavarian},
  \bibinfo{person}{Clemens Winter}, \bibinfo{person}{Philippe Tillet},
  \bibinfo{person}{Felipe~Petroski Such}, \bibinfo{person}{Dave Cummings},
  \bibinfo{person}{Matthias Plappert}, \bibinfo{person}{Fotios Chantzis},
  \bibinfo{person}{Elizabeth Barnes}, \bibinfo{person}{Ariel Herbert-Voss},
  \bibinfo{person}{William~Hebgen Guss}, \bibinfo{person}{Alex Nichol},
  \bibinfo{person}{Alex Paino}, \bibinfo{person}{Nikolas Tezak},
  \bibinfo{person}{Jie Tang}, \bibinfo{person}{Igor Babuschkin},
  \bibinfo{person}{Suchir Balaji}, \bibinfo{person}{Shantanu Jain},
  \bibinfo{person}{William Saunders}, \bibinfo{person}{Christopher Hesse},
  \bibinfo{person}{Andrew~N. Carr}, \bibinfo{person}{Jan Leike},
  \bibinfo{person}{Josh Achiam}, \bibinfo{person}{Vedant Misra},
  \bibinfo{person}{Evan Morikawa}, \bibinfo{person}{Alec Radford},
  \bibinfo{person}{Matthew Knight}, \bibinfo{person}{Miles Brundage},
  \bibinfo{person}{Mira Murati}, \bibinfo{person}{Katie Mayer},
  \bibinfo{person}{Peter Welinder}, \bibinfo{person}{Bob McGrew},
  \bibinfo{person}{Dario Amodei}, \bibinfo{person}{Sam McCandlish},
  \bibinfo{person}{Ilya Sutskever}, {and} \bibinfo{person}{Wojciech Zaremba}.}
  \bibinfo{year}{2021}\natexlab{}.
\newblock \bibinfo{title}{Evaluating Large Language Models Trained on Code}.
\newblock
\newblock
\urldef\tempurl%
\url{https://doi.org/10.48550/ARXIV.2107.03374}
\showDOI{\tempurl}


\bibitem[\protect\citeauthoryear{D'Antoni, Samanta, and Singh}{D'Antoni
  et~al\mbox{.}}{2016}]%
        {QLOSE}
\bibfield{author}{\bibinfo{person}{Loris D'Antoni}, \bibinfo{person}{Roopsha
  Samanta}, {and} \bibinfo{person}{Rishabh Singh}.}
  \bibinfo{year}{2016}\natexlab{}.
\newblock \showarticletitle{Qlose: Program Repair with Quantitative
  Objectives}. In \bibinfo{booktitle}{\emph{Computer Aided Verification - 28th
  International Conference, {CAV} 2016, Toronto, ON, Canada, July 17-23, 2016,
  Proceedings, Part {II}}} \emph{(\bibinfo{series}{Lecture Notes in Computer
  Science}, Vol.~\bibinfo{volume}{9780})},
  \bibfield{editor}{\bibinfo{person}{Swarat Chaudhuri} {and}
  \bibinfo{person}{Azadeh Farzan}} (Eds.). \bibinfo{publisher}{Springer},
  \bibinfo{pages}{383--401}.
\newblock
\urldef\tempurl%
\url{https://doi.org/10.1007/978-3-319-41540-6\_21}
\showDOI{\tempurl}


\bibitem[\protect\citeauthoryear{Goues, Forrest, and Weimer}{Goues
  et~al\mbox{.}}{2013}]%
        {APRGoues13}
\bibfield{author}{\bibinfo{person}{Claire Goues}, \bibinfo{person}{Stephanie
  Forrest}, {and} \bibinfo{person}{Westley Weimer}.}
  \bibinfo{year}{2013}\natexlab{}.
\newblock \showarticletitle{Current Challenges in Automatic Software Repair}.
\newblock \bibinfo{journal}{\emph{Software Quality Journal}}
  \bibinfo{volume}{21}, \bibinfo{number}{3} (\bibinfo{date}{sep}
  \bibinfo{year}{2013}), \bibinfo{pages}{421–443}.
\newblock
\showISSN{0963-9314}
\urldef\tempurl%
\url{https://doi.org/10.1007/s11219-013-9208-0}
\showDOI{\tempurl}


\bibitem[\protect\citeauthoryear{Goues, Pradel, and Roychoudhury}{Goues
  et~al\mbox{.}}{2019}]%
        {APRGoues19}
\bibfield{author}{\bibinfo{person}{Claire~Le Goues}, \bibinfo{person}{Michael
  Pradel}, {and} \bibinfo{person}{Abhik Roychoudhury}.}
  \bibinfo{year}{2019}\natexlab{}.
\newblock \showarticletitle{Automated Program Repair}.
\newblock \bibinfo{journal}{\emph{Commun. ACM}} \bibinfo{volume}{62},
  \bibinfo{number}{12} (\bibinfo{date}{nov} \bibinfo{year}{2019}),
  \bibinfo{pages}{56–65}.
\newblock
\showISSN{0001-0782}
\urldef\tempurl%
\url{https://doi.org/10.1145/3318162}
\showDOI{\tempurl}


\bibitem[\protect\citeauthoryear{Gulwani, Radicek, and Zuleger}{Gulwani
  et~al\mbox{.}}{2018}]%
        {clara}
\bibfield{author}{\bibinfo{person}{Sumit Gulwani}, \bibinfo{person}{Ivan
  Radicek}, {and} \bibinfo{person}{Florian Zuleger}.}
  \bibinfo{year}{2018}\natexlab{}.
\newblock \showarticletitle{Automated Clustering and Program Repair for
  Introductory Programming Assignments}.
\newblock \bibinfo{journal}{\emph{SIGPLAN Not.}} \bibinfo{volume}{53},
  \bibinfo{number}{4} (\bibinfo{date}{June} \bibinfo{year}{2018}),
  \bibinfo{pages}{465–480}.
\newblock
\showISSN{0362-1340}
\urldef\tempurl%
\url{https://doi.org/10.1145/3296979.3192387}
\showDOI{\tempurl}


\bibitem[\protect\citeauthoryear{Hendrycks, Basart, Kadavath, Mazeika, Arora,
  Guo, Burns, Puranik, He, Song, and Steinhardt}{Hendrycks
  et~al\mbox{.}}{2021}]%
        {APPS}
\bibfield{author}{\bibinfo{person}{Dan Hendrycks}, \bibinfo{person}{Steven
  Basart}, \bibinfo{person}{Saurav Kadavath}, \bibinfo{person}{Mantas Mazeika},
  \bibinfo{person}{Akul Arora}, \bibinfo{person}{Ethan Guo},
  \bibinfo{person}{Collin Burns}, \bibinfo{person}{Samir Puranik},
  \bibinfo{person}{Horace He}, \bibinfo{person}{Dawn Song}, {and}
  \bibinfo{person}{Jacob Steinhardt}.} \bibinfo{year}{2021}\natexlab{}.
\newblock \showarticletitle{Measuring Coding Challenge Competence With {APPS}}.
\newblock \bibinfo{journal}{\emph{CoRR}}  \bibinfo{volume}{abs/2105.09938}
  (\bibinfo{year}{2021}).
\newblock
\showeprint[arXiv]{2105.09938}
\urldef\tempurl%
\url{https://arxiv.org/abs/2105.09938}
\showURL{%
\tempurl}


\bibitem[\protect\citeauthoryear{{Hu}, {Ahmed}, {Mechtaev}, {Leong}, and
  {Roychoudhury}}{{Hu} et~al\mbox{.}}{2019}]%
        {Refactory}
\bibfield{author}{\bibinfo{person}{Y. {Hu}}, \bibinfo{person}{U.~Z. {Ahmed}},
  \bibinfo{person}{S. {Mechtaev}}, \bibinfo{person}{B. {Leong}}, {and}
  \bibinfo{person}{A. {Roychoudhury}}.} \bibinfo{year}{2019}\natexlab{}.
\newblock \showarticletitle{Re-Factoring Based Program Repair Applied to
  Programming Assignments}. In \bibinfo{booktitle}{\emph{2019 34th IEEE/ACM
  International Conference on Automated Software Engineering (ASE)}}.
  \bibinfo{pages}{388--398}.
\newblock


\bibitem[\protect\citeauthoryear{Kaleeswaran, Santhiar, Kanade, and
  Gulwani}{Kaleeswaran et~al\mbox{.}}{2016}]%
        {CoderAssist}
\bibfield{author}{\bibinfo{person}{Shalini Kaleeswaran},
  \bibinfo{person}{Anirudh Santhiar}, \bibinfo{person}{Aditya Kanade}, {and}
  \bibinfo{person}{Sumit Gulwani}.} \bibinfo{year}{2016}\natexlab{}.
\newblock \showarticletitle{Semi-Supervised Verified Feedback Generation}. In
  \bibinfo{booktitle}{\emph{Proceedings of the 2016 24th ACM SIGSOFT
  International Symposium on Foundations of Software Engineering}} (Seattle,
  WA, USA) \emph{(\bibinfo{series}{FSE 2016})}. \bibinfo{publisher}{Association
  for Computing Machinery}, \bibinfo{address}{New York, NY, USA},
  \bibinfo{pages}{739–750}.
\newblock
\showISBNx{9781450342186}
\urldef\tempurl%
\url{https://doi.org/10.1145/2950290.2950363}
\showDOI{\tempurl}


\bibitem[\protect\citeauthoryear{Kim, Nam, Song, and Kim}{Kim
  et~al\mbox{.}}{2013}]%
        {PAR}
\bibfield{author}{\bibinfo{person}{Dongsun Kim}, \bibinfo{person}{Jaechang
  Nam}, \bibinfo{person}{Jaewoo Song}, {and} \bibinfo{person}{Sunghun Kim}.}
  \bibinfo{year}{2013}\natexlab{}.
\newblock \showarticletitle{Automatic Patch Generation Learned from
  Human-Written Patches}. In \bibinfo{booktitle}{\emph{Proceedings of the 2013
  International Conference on Software Engineering}} (San Francisco, CA, USA)
  \emph{(\bibinfo{series}{ICSE '13})}. \bibinfo{publisher}{IEEE Press},
  \bibinfo{pages}{802–811}.
\newblock
\showISBNx{9781467330763}


\bibitem[\protect\citeauthoryear{Laaksonen}{Laaksonen}{2020}]%
        {CompetitiveProgrammingGuide}
\bibfield{author}{\bibinfo{person}{Antti Laaksonen}.}
  \bibinfo{year}{2020}\natexlab{}.
\newblock \bibinfo{booktitle}{\emph{Guide to Competitive Programming - Learning
  and Improving Algorithms Through Contests, Second Edition}}.
\newblock \bibinfo{publisher}{Springer}.
\newblock
\showISBNx{978-3-030-39356-4}
\urldef\tempurl%
\url{https://doi.org/10.1007/978-3-030-39357-1}
\showDOI{\tempurl}


\bibitem[\protect\citeauthoryear{Le~Goues, Nguyen, Forrest, and
  Weimer}{Le~Goues et~al\mbox{.}}{2012}]%
        {GenProg}
\bibfield{author}{\bibinfo{person}{Claire Le~Goues}, \bibinfo{person}{ThanhVu
  Nguyen}, \bibinfo{person}{Stephanie Forrest}, {and} \bibinfo{person}{Westley
  Weimer}.} \bibinfo{year}{2012}\natexlab{}.
\newblock \showarticletitle{GenProg: A Generic Method for Automatic Software
  Repair}.
\newblock \bibinfo{journal}{\emph{IEEE Transactions on Software Engineering}}
  \bibinfo{volume}{38}, \bibinfo{number}{1} (\bibinfo{year}{2012}),
  \bibinfo{pages}{54--72}.
\newblock
\urldef\tempurl%
\url{https://doi.org/10.1109/TSE.2011.104}
\showDOI{\tempurl}


\bibitem[\protect\citeauthoryear{Lee, Song, So, and Oh}{Lee
  et~al\mbox{.}}{2018}]%
        {FixML}
\bibfield{author}{\bibinfo{person}{Junho Lee}, \bibinfo{person}{Dowon Song},
  \bibinfo{person}{Sunbeom So}, {and} \bibinfo{person}{Hakjoo Oh}.}
  \bibinfo{year}{2018}\natexlab{}.
\newblock \showarticletitle{Automatic Diagnosis and Correction of Logical
  Errors for Functional Programming Assignments}.
\newblock \bibinfo{journal}{\emph{Proc. ACM Program. Lang.}}
  \bibinfo{volume}{2}, \bibinfo{number}{OOPSLA}, Article
  \bibinfo{articleno}{158} (\bibinfo{date}{oct} \bibinfo{year}{2018}),
  \bibinfo{numpages}{30}~pages.
\newblock
\urldef\tempurl%
\url{https://doi.org/10.1145/3276528}
\showDOI{\tempurl}


\bibitem[\protect\citeauthoryear{Li, Choi, Chung, Kushman, Schrittwieser,
  Leblond, Eccles, Keeling, Gimeno, Dal~Lago, Hubert, Choy, de~Masson~d'Autume,
  Babuschkin, Chen, Huang, Welbl, Gowal, Cherepanov, Molloy, Mankowitz,
  Sutherland~Robson, Kohli, de~Freitas, Kavukcuoglu, and Vinyals}{Li
  et~al\mbox{.}}{2022}]%
        {alphacode}
\bibfield{author}{\bibinfo{person}{Yujia Li}, \bibinfo{person}{David Choi},
  \bibinfo{person}{Junyoung Chung}, \bibinfo{person}{Nate Kushman},
  \bibinfo{person}{Julian Schrittwieser}, \bibinfo{person}{Rémi Leblond},
  \bibinfo{person}{Tom Eccles}, \bibinfo{person}{James Keeling},
  \bibinfo{person}{Felix Gimeno}, \bibinfo{person}{Agustin Dal~Lago},
  \bibinfo{person}{Thomas Hubert}, \bibinfo{person}{Peter Choy},
  \bibinfo{person}{Cyprien de Masson~d'Autume}, \bibinfo{person}{Igor
  Babuschkin}, \bibinfo{person}{Xinyun Chen}, \bibinfo{person}{Po-Sen Huang},
  \bibinfo{person}{Johannes Welbl}, \bibinfo{person}{Sven Gowal},
  \bibinfo{person}{Alexey Cherepanov}, \bibinfo{person}{James Molloy},
  \bibinfo{person}{Daniel Mankowitz}, \bibinfo{person}{Esme Sutherland~Robson},
  \bibinfo{person}{Pushmeet Kohli}, \bibinfo{person}{Nando de Freitas},
  \bibinfo{person}{Koray Kavukcuoglu}, {and} \bibinfo{person}{Oriol Vinyals}.}
  \bibinfo{year}{2022}\natexlab{}.
\newblock \bibinfo{title}{Competition-Level Code Generation with AlphaCode}.
\newblock
\newblock


\bibitem[\protect\citeauthoryear{Long, Amidon, and Rinard}{Long
  et~al\mbox{.}}{2017}]%
        {genesis}
\bibfield{author}{\bibinfo{person}{Fan Long}, \bibinfo{person}{Peter Amidon},
  {and} \bibinfo{person}{Martin Rinard}.} \bibinfo{year}{2017}\natexlab{}.
\newblock \showarticletitle{Automatic Inference of Code Transforms for Patch
  Generation}. In \bibinfo{booktitle}{\emph{Proceedings of the 2017 11th Joint
  Meeting on Foundations of Software Engineering}} (Paderborn, Germany)
  \emph{(\bibinfo{series}{ESEC/FSE 2017})}. \bibinfo{publisher}{Association for
  Computing Machinery}, \bibinfo{address}{New York, NY, USA},
  \bibinfo{pages}{727–739}.
\newblock
\showISBNx{9781450351058}
\urldef\tempurl%
\url{https://doi.org/10.1145/3106237.3106253}
\showDOI{\tempurl}


\bibitem[\protect\citeauthoryear{Long and Rinard}{Long and Rinard}{2016}]%
        {prophet}
\bibfield{author}{\bibinfo{person}{Fan Long} {and} \bibinfo{person}{Martin
  Rinard}.} \bibinfo{year}{2016}\natexlab{}.
\newblock \showarticletitle{Automatic Patch Generation by Learning Correct
  Code}. In \bibinfo{booktitle}{\emph{Proceedings of the 43rd Annual ACM
  SIGPLAN-SIGACT Symposium on Principles of Programming Languages}} (St.
  Petersburg, FL, USA) \emph{(\bibinfo{series}{POPL '16})}.
  \bibinfo{publisher}{Association for Computing Machinery},
  \bibinfo{address}{New York, NY, USA}, \bibinfo{pages}{298–312}.
\newblock
\showISBNx{9781450335492}
\urldef\tempurl%
\url{https://doi.org/10.1145/2837614.2837617}
\showDOI{\tempurl}


\bibitem[\protect\citeauthoryear{Lu, Meng, and Li}{Lu et~al\mbox{.}}{2021}]%
        {FAPR}
\bibfield{author}{\bibinfo{person}{Yunlong Lu}, \bibinfo{person}{Na Meng},
  {and} \bibinfo{person}{Wenxin Li}.} \bibinfo{year}{2021}\natexlab{}.
\newblock \showarticletitle{{FAPR:} Fast and Accurate Program Repair for
  Introductory Programming Courses}.
\newblock \bibinfo{journal}{\emph{CoRR}}  \bibinfo{volume}{abs/2107.06550}
  (\bibinfo{year}{2021}).
\newblock
\showeprint[arXiv]{2107.06550}
\urldef\tempurl%
\url{https://arxiv.org/abs/2107.06550}
\showURL{%
\tempurl}


\bibitem[\protect\citeauthoryear{Mechtaev, Nguyen, Noller, Grunske, and
  Roychoudhury}{Mechtaev et~al\mbox{.}}{2018}]%
        {MechtaevNNGR18}
\bibfield{author}{\bibinfo{person}{Sergey Mechtaev},
  \bibinfo{person}{Manh{-}Dung Nguyen}, \bibinfo{person}{Yannic Noller},
  \bibinfo{person}{Lars Grunske}, {and} \bibinfo{person}{Abhik Roychoudhury}.}
  \bibinfo{year}{2018}\natexlab{}.
\newblock \showarticletitle{Semantic program repair using a reference
  implementation}. In \bibinfo{booktitle}{\emph{Proceedings of the 40th
  International Conference on Software Engineering, {ICSE} 2018, Gothenburg,
  Sweden, May 27 - June 03, 2018}}, \bibfield{editor}{\bibinfo{person}{Michel
  Chaudron}, \bibinfo{person}{Ivica Crnkovic}, \bibinfo{person}{Marsha
  Chechik}, {and} \bibinfo{person}{Mark Harman}} (Eds.).
  \bibinfo{publisher}{{ACM}}, \bibinfo{pages}{129--139}.
\newblock
\urldef\tempurl%
\url{https://doi.org/10.1145/3180155.3180247}
\showDOI{\tempurl}


\bibitem[\protect\citeauthoryear{Mechtaev, Yi, and Roychoudhury}{Mechtaev
  et~al\mbox{.}}{2016}]%
        {Angelix}
\bibfield{author}{\bibinfo{person}{Sergey Mechtaev}, \bibinfo{person}{Jooyong
  Yi}, {and} \bibinfo{person}{Abhik Roychoudhury}.}
  \bibinfo{year}{2016}\natexlab{}.
\newblock \showarticletitle{Angelix: Scalable Multiline Program Patch Synthesis
  via Symbolic Analysis}. In \bibinfo{booktitle}{\emph{2016 IEEE/ACM 38th
  International Conference on Software Engineering (ICSE)}}.
  \bibinfo{pages}{691--701}.
\newblock
\urldef\tempurl%
\url{https://doi.org/10.1145/2884781.2884807}
\showDOI{\tempurl}


\bibitem[\protect\citeauthoryear{Nguyen and Xu}{Nguyen and Xu}{2013}]%
        {Cachetor}
\bibfield{author}{\bibinfo{person}{Khanh Nguyen} {and} \bibinfo{person}{Guoqing
  Xu}.} \bibinfo{year}{2013}\natexlab{}.
\newblock \showarticletitle{Cachetor: Detecting Cacheable Data to Remove
  Bloat}. In \bibinfo{booktitle}{\emph{Proceedings of the 2013 9th Joint
  Meeting on Foundations of Software Engineering}} (Saint Petersburg, Russia)
  \emph{(\bibinfo{series}{ESEC/FSE 2013})}. \bibinfo{publisher}{Association for
  Computing Machinery}, \bibinfo{address}{New York, NY, USA},
  \bibinfo{pages}{268–278}.
\newblock
\showISBNx{9781450322379}
\urldef\tempurl%
\url{https://doi.org/10.1145/2491411.2491416}
\showDOI{\tempurl}


\bibitem[\protect\citeauthoryear{Nistor, Chang, Radoi, and Lu}{Nistor
  et~al\mbox{.}}{2015}]%
        {Caramel}
\bibfield{author}{\bibinfo{person}{Adrian Nistor}, \bibinfo{person}{Po-Chun
  Chang}, \bibinfo{person}{Cosmin Radoi}, {and} \bibinfo{person}{Shan Lu}.}
  \bibinfo{year}{2015}\natexlab{}.
\newblock \showarticletitle{Caramel: Detecting and Fixing Performance Problems
  That Have Non-Intrusive Fixes}. In \bibinfo{booktitle}{\emph{Proceedings of
  the 37th International Conference on Software Engineering - Volume 1}}
  (Florence, Italy) \emph{(\bibinfo{series}{ICSE '15})}.
  \bibinfo{publisher}{IEEE Press}, \bibinfo{pages}{902–912}.
\newblock
\showISBNx{9781479919345}


\bibitem[\protect\citeauthoryear{Perry, Kim, Samanta, and Zhang}{Perry
  et~al\mbox{.}}{2019}]%
        {SemCluster}
\bibfield{author}{\bibinfo{person}{David~M. Perry}, \bibinfo{person}{Dohyeong
  Kim}, \bibinfo{person}{Roopsha Samanta}, {and} \bibinfo{person}{Xiangyu
  Zhang}.} \bibinfo{year}{2019}\natexlab{}.
\newblock \showarticletitle{SemCluster: Clustering of Imperative Programming
  Assignments Based on Quantitative Semantic Features}. In
  \bibinfo{booktitle}{\emph{Proceedings of the 40th ACM SIGPLAN Conference on
  Programming Language Design and Implementation}} (Phoenix, AZ, USA)
  \emph{(\bibinfo{series}{PLDI 2019})}. \bibinfo{publisher}{Association for
  Computing Machinery}, \bibinfo{address}{New York, NY, USA},
  \bibinfo{pages}{860–873}.
\newblock
\showISBNx{9781450367127}
\urldef\tempurl%
\url{https://doi.org/10.1145/3314221.3314629}
\showDOI{\tempurl}


\bibitem[\protect\citeauthoryear{Pu, Narasimhan, Solar-Lezama, and Barzilay}{Pu
  et~al\mbox{.}}{2016}]%
        {SKP}
\bibfield{author}{\bibinfo{person}{Yewen Pu}, \bibinfo{person}{Karthik
  Narasimhan}, \bibinfo{person}{Armando Solar-Lezama}, {and}
  \bibinfo{person}{Regina Barzilay}.} \bibinfo{year}{2016}\natexlab{}.
\newblock \showarticletitle{Sk\_p: A Neural Program Corrector for MOOCs}. In
  \bibinfo{booktitle}{\emph{Companion Proceedings of the 2016 ACM SIGPLAN
  International Conference on Systems, Programming, Languages and Applications:
  Software for Humanity}} (Amsterdam, Netherlands)
  \emph{(\bibinfo{series}{SPLASH Companion 2016})}.
  \bibinfo{publisher}{Association for Computing Machinery},
  \bibinfo{address}{New York, NY, USA}, \bibinfo{pages}{39–40}.
\newblock
\showISBNx{9781450344371}
\urldef\tempurl%
\url{https://doi.org/10.1145/2984043.2989222}
\showDOI{\tempurl}


\bibitem[\protect\citeauthoryear{Puri, Kung, Janssen, Zhang, Domeniconi,
  Zolotov, Dolby, Chen, Choudhury, Decker, Thost, Buratti, Pujar, and
  Finkler}{Puri et~al\mbox{.}}{2021}]%
        {Puri}
\bibfield{author}{\bibinfo{person}{Ruchir Puri}, \bibinfo{person}{David~S.
  Kung}, \bibinfo{person}{Geert Janssen}, \bibinfo{person}{Wei Zhang},
  \bibinfo{person}{Giacomo Domeniconi}, \bibinfo{person}{Vladimir Zolotov},
  \bibinfo{person}{Julian Dolby}, \bibinfo{person}{Jie Chen},
  \bibinfo{person}{Mihir~R. Choudhury}, \bibinfo{person}{Lindsey Decker},
  \bibinfo{person}{Veronika Thost}, \bibinfo{person}{Luca Buratti},
  \bibinfo{person}{Saurabh Pujar}, {and} \bibinfo{person}{Ulrich Finkler}.}
  \bibinfo{year}{2021}\natexlab{}.
\newblock \showarticletitle{Project CodeNet: {A} Large-Scale {AI} for Code
  Dataset for Learning a Diversity of Coding Tasks}.
\newblock \bibinfo{journal}{\emph{CoRR}}  \bibinfo{volume}{abs/2105.12655}
  (\bibinfo{year}{2021}).
\newblock
\showeprint[arXiv]{2105.12655}
\urldef\tempurl%
\url{https://arxiv.org/abs/2105.12655}
\showURL{%
\tempurl}


\bibitem[\protect\citeauthoryear{Qi, Mao, Lei, Dai, and Wang}{Qi
  et~al\mbox{.}}{2014}]%
        {RandomQi}
\bibfield{author}{\bibinfo{person}{Yuhua Qi}, \bibinfo{person}{Xiaoguang Mao},
  \bibinfo{person}{Yan Lei}, \bibinfo{person}{Ziying Dai}, {and}
  \bibinfo{person}{Chengsong Wang}.} \bibinfo{year}{2014}\natexlab{}.
\newblock \showarticletitle{The Strength of Random Search on Automated Program
  Repair}. In \bibinfo{booktitle}{\emph{Proceedings of the 36th International
  Conference on Software Engineering}} (Hyderabad, India)
  \emph{(\bibinfo{series}{ICSE 2014})}. \bibinfo{publisher}{Association for
  Computing Machinery}, \bibinfo{address}{New York, NY, USA},
  \bibinfo{pages}{254–265}.
\newblock
\showISBNx{9781450327565}
\urldef\tempurl%
\url{https://doi.org/10.1145/2568225.2568254}
\showDOI{\tempurl}


\bibitem[\protect\citeauthoryear{{Rolim}, {Soares}, {D'Antoni}, {Polozov},
  {Gulwani}, {Gheyi}, {Suzuki}, and {Hartmann}}{{Rolim} et~al\mbox{.}}{2017}]%
        {Refazer}
\bibfield{author}{\bibinfo{person}{R. {Rolim}}, \bibinfo{person}{G. {Soares}},
  \bibinfo{person}{L. {D'Antoni}}, \bibinfo{person}{O. {Polozov}},
  \bibinfo{person}{S. {Gulwani}}, \bibinfo{person}{R. {Gheyi}},
  \bibinfo{person}{R. {Suzuki}}, {and} \bibinfo{person}{B. {Hartmann}}.}
  \bibinfo{year}{2017}\natexlab{}.
\newblock \showarticletitle{Learning Syntactic Program Transformations from
  Examples}. In \bibinfo{booktitle}{\emph{2017 IEEE/ACM 39th International
  Conference on Software Engineering (ICSE)}}. \bibinfo{pages}{404--415}.
\newblock


\bibitem[\protect\citeauthoryear{Singh, Gulwani, and Solar-Lezama}{Singh
  et~al\mbox{.}}{2013}]%
        {Autograder}
\bibfield{author}{\bibinfo{person}{Rishabh Singh}, \bibinfo{person}{Sumit
  Gulwani}, {and} \bibinfo{person}{Armando Solar-Lezama}.}
  \bibinfo{year}{2013}\natexlab{}.
\newblock \showarticletitle{Automated Feedback Generation for Introductory
  Programming Assignments}. In \bibinfo{booktitle}{\emph{Proceedings of the
  34th ACM SIGPLAN Conference on Programming Language Design and
  Implementation}} (Seattle, Washington, USA) \emph{(\bibinfo{series}{PLDI
  13})}. \bibinfo{publisher}{Association for Computing Machinery},
  \bibinfo{address}{New York, NY, USA}, \bibinfo{pages}{15–26}.
\newblock
\showISBNx{9781450320146}
\urldef\tempurl%
\url{https://doi.org/10.1145/2491956.2462195}
\showDOI{\tempurl}


\bibitem[\protect\citeauthoryear{Song, Lee, and Oh}{Song et~al\mbox{.}}{2021}]%
        {Cafe}
\bibfield{author}{\bibinfo{person}{Dowon Song}, \bibinfo{person}{Woosuk Lee},
  {and} \bibinfo{person}{Hakjoo Oh}.} \bibinfo{year}{2021}\natexlab{}.
\newblock \bibinfo{booktitle}{\emph{Context-Aware and Data-Driven Feedback
  Generation for Programming Assignments}}.
\newblock \bibinfo{publisher}{Association for Computing Machinery},
  \bibinfo{address}{New York, NY, USA}, \bibinfo{pages}{328–340}.
\newblock
\showISBNx{9781450385626}
\urldef\tempurl%
\url{https://doi.org/10.1145/3468264.3468598}
\showURL{%
\tempurl}


\bibitem[\protect\citeauthoryear{Wang, Singh, and Su}{Wang
  et~al\mbox{.}}{2018a}]%
        {SAR}
\bibfield{author}{\bibinfo{person}{Ke Wang}, \bibinfo{person}{Rishabh Singh},
  {and} \bibinfo{person}{Zhendong Su}.} \bibinfo{year}{2018}\natexlab{a}.
\newblock \showarticletitle{Search, Align, and Repair: Data-Driven Feedback
  Generation for Introductory Programming Exercises}
  \emph{(\bibinfo{series}{PLDI 2018})}. \bibinfo{publisher}{Association for
  Computing Machinery}, \bibinfo{address}{New York, NY, USA},
  \bibinfo{pages}{481–495}.
\newblock
\showISBNx{9781450356985}
\urldef\tempurl%
\url{https://doi.org/10.1145/3192366.3192384}
\showDOI{\tempurl}


\bibitem[\protect\citeauthoryear{Wang, Su, and Singh}{Wang
  et~al\mbox{.}}{2018b}]%
        {wang2018dynamic}
\bibfield{author}{\bibinfo{person}{Ke Wang}, \bibinfo{person}{Zhendong Su},
  {and} \bibinfo{person}{Rishabh Singh}.} \bibinfo{year}{2018}\natexlab{b}.
\newblock \showarticletitle{Dynamic Neural Program Embeddings for Program
  Repair}. In \bibinfo{booktitle}{\emph{International Conference on Learning
  Representations}}.
\newblock
\urldef\tempurl%
\url{https://openreview.net/forum?id=BJuWrGW0Z}
\showURL{%
\tempurl}


\bibitem[\protect\citeauthoryear{Yasunaga and Liang}{Yasunaga and
  Liang}{2021}]%
        {BIFI}
\bibfield{author}{\bibinfo{person}{Michihiro Yasunaga} {and}
  \bibinfo{person}{Percy Liang}.} \bibinfo{year}{2021}\natexlab{}.
\newblock \showarticletitle{Break-It-Fix-It: Unsupervised Learning for Program
  Repair}.
\newblock \bibinfo{journal}{\emph{CoRR}}  \bibinfo{volume}{abs/2106.06600}
  (\bibinfo{year}{2021}).
\newblock
\showeprint[arXiv]{2106.06600}
\urldef\tempurl%
\url{https://arxiv.org/abs/2106.06600}
\showURL{%
\tempurl}


\bibitem[\protect\citeauthoryear{Yi, Ahmed, Karkare, Tan, and Roychoudhury}{Yi
  et~al\mbox{.}}{2017}]%
        {YiFSE17}
\bibfield{author}{\bibinfo{person}{Jooyong Yi}, \bibinfo{person}{Umair~Z.
  Ahmed}, \bibinfo{person}{Amey Karkare}, \bibinfo{person}{Shin~Hwei Tan},
  {and} \bibinfo{person}{Abhik Roychoudhury}.} \bibinfo{year}{2017}\natexlab{}.
\newblock \showarticletitle{A Feasibility Study of Using Automated Program
  Repair for Introductory Programming Assignments}. In
  \bibinfo{booktitle}{\emph{Proceedings of the 2017 11th Joint Meeting on
  Foundations of Software Engineering}} (Paderborn, Germany)
  \emph{(\bibinfo{series}{ESEC/FSE 2017})}. \bibinfo{publisher}{Association for
  Computing Machinery}, \bibinfo{address}{New York, NY, USA},
  \bibinfo{pages}{740–751}.
\newblock
\showISBNx{9781450351058}
\urldef\tempurl%
\url{https://doi.org/10.1145/3106237.3106262}
\showDOI{\tempurl}


\bibitem[\protect\citeauthoryear{Zavershynskyi, Skidanov, and
  Polosukhin}{Zavershynskyi et~al\mbox{.}}{2018}]%
        {NAPS}
\bibfield{author}{\bibinfo{person}{Maksym Zavershynskyi},
  \bibinfo{person}{Alexander Skidanov}, {and} \bibinfo{person}{Illia
  Polosukhin}.} \bibinfo{year}{2018}\natexlab{}.
\newblock \showarticletitle{{NAPS:} Natural Program Synthesis Dataset}.
\newblock \bibinfo{journal}{\emph{CoRR}}  \bibinfo{volume}{abs/1807.03168}
  (\bibinfo{year}{2018}).
\newblock
\showeprint[arXiv]{1807.03168}
\urldef\tempurl%
\url{http://arxiv.org/abs/1807.03168}
\showURL{%
\tempurl}


\bibitem[\protect\citeauthoryear{Zhang and Shasha}{Zhang and Shasha}{1989a}]%
        {zss}
\bibfield{author}{\bibinfo{person}{Kaizhong Zhang} {and}
  \bibinfo{person}{Dennis Shasha}.} \bibinfo{year}{1989}\natexlab{a}.
\newblock \showarticletitle{Simple Fast Algorithms for the Editing Distance
  between Trees and Related Problems}.
\newblock \bibinfo{journal}{\emph{SIAM J. Comput.}} \bibinfo{volume}{18},
  \bibinfo{number}{6} (\bibinfo{year}{1989}), \bibinfo{pages}{1245--1262}.
\newblock
\urldef\tempurl%
\url{https://doi.org/10.1137/0218082}
\showDOI{\tempurl}


\bibitem[\protect\citeauthoryear{Zhang and Shasha}{Zhang and Shasha}{1989b}]%
        {zhangshasha}
\bibfield{author}{\bibinfo{person}{Kaizhong Zhang} {and}
  \bibinfo{person}{Dennis~E. Shasha}.} \bibinfo{year}{1989}\natexlab{b}.
\newblock \showarticletitle{Simple Fast Algorithms for the Editing Distance
  Between Trees and Related Problems}.
\newblock \bibinfo{journal}{\emph{{SIAM} J. Comput.}} \bibinfo{volume}{18},
  \bibinfo{number}{6} (\bibinfo{year}{1989}), \bibinfo{pages}{1245--1262}.
\newblock
\urldef\tempurl%
\url{https://doi.org/10.1137/0218082}
\showDOI{\tempurl}


\end{thebibliography}
%\bibliography{os}
\end{document}